\documentclass[twocolumn, twocolappendix]{aastex63}

\usepackage{graphicx}	% Including figure files
\usepackage{amsmath}	% Advanced maths commands
\usepackage{amssymb}	% Extra maths symbols
\usepackage{multirow}   % allow entries in tables to span multiple columns or rows

\received{tbd}
\revised{tbd}
\accepted{tbd}
\submitjournal{AJ}

\shorttitle{MYSST I}
\shortauthors{Ksoll et al.}

\begin{document}

%%%%%%%%%% TITLE %%%%%%%%%%%
\title{Measuring Young Stars in Space and Time - I. The Photometric Catalog and Extinction Properties of N44}

%%%%%%%%%%%%%%%%%%%%%%%%%%%%
%%%%%%%%% AUTHORS %%%%%%%%%%
%%%%%%%%%%%%%%%%%%%%%%%%%%%%

%% The \author command is the same as before except it now takes an optional
%% argument which is the 16 digit ORCID. The syntax is:
%% \author[xxxx-xxxx-xxxx-xxxx]{Author Name}
%% Use \affiliation for affiliation information. When a duplicate is found its index will be the same as its previous entry.
%%
%% Use \email to set provide email addresses. Each \email will appear on its
%% own line so you can put multiple email address in one \email call. A new
%% \correspondingauthor command is available in V6.3 to identify the
%% corresponding author of the manuscript. 
%%
%% While authors can be grouped inside the same \author and \affiliation
%% commands it is better to have a single author for each. This allows for
%% one to exploit all the new benefits and should make book-keeping easier.
%%

\correspondingauthor{Victor F. Ksoll}
\email{v.ksoll@stud.uni-heidelberg.de}

\author[0000-0002-0294-799X]{Victor F.\ Ksoll}
\affiliation{Universit\"{a}t Heidelberg, Zentrum f\"{u}r Astronomie, Institut f\"{u}r Theoretische Astrophysik,\\ Albert-Ueberle-Str. 2, 69120 Heidelberg, Germany}
\affiliation{Universit\"{a}t Heidelberg, Interdisziplin\"{a}res Zentrum f\"{u}r Wissenschaftliches Rechnen,\\ Im Neuenheimer Feld 205, 69120 Heidelberg, Germany}

\author[0000-0002-2763-0075]{Dimitrios Gouliermis}
\affiliation{Universit\"{a}t Heidelberg, Zentrum f\"{u}r Astronomie, Institut f\"{u}r Theoretische Astrophysik,\\ Albert-Ueberle-Str. 2, 69120 Heidelberg, Germany}
\affiliation{Max Planck Institute for Astronomy, K\"{o}nigstuhl\,17, 69117 Heidelberg, Germany}

\author[0000-0003-2954-7643]{Elena Sabbi}
\affiliation{Space Telescope Science Institute, 3700 San Martin Drive, Baltimore, MD 21218, USA}

\author{Jenna E.\ Ryon}
\affiliation{Space Telescope Science Institute, 3700 San Martin Drive, Baltimore, MD 21218, USA}

\author[0000-0002-9573-3199]{Massimo Robberto}
\affiliation{Space Telescope Science Institute, 3700 San Martin Drive, Baltimore, MD 21218, USA}
\affiliation{Johns Hopkins University, 3400 N. Charles Street, Baltimore, MD 21218, USA}

\author[0000-0002-5581-2896]{Mario Gennaro}
\affiliation{Space Telescope Science Institute, 3700 San Martin Drive, Baltimore, MD 21218, USA}
\affiliation{Johns Hopkins University, 3400 N. Charles Street, Baltimore, MD 21218, USA}

\author[0000-0002-0560-3172]{Ralf S.\ Klessen}
\affiliation{Universit\"{a}t Heidelberg, Zentrum f\"{u}r Astronomie, Institut f\"{u}r Theoretische Astrophysik,\\  Albert-Ueberle-Str. 2, 69120 Heidelberg, Germany}
\affiliation{Universit\"{a}t Heidelberg, Interdisziplin\"{a}res Zentrum f\"{u}r Wissenschaftliches Rechnen,\\ Im Neuenheimer Feld 205, 69120 Heidelberg, Germany}

\author[0000-0001-6036-1287]{Ullrich Koethe}
\affiliation{Universit\"{a}t Heidelberg, Heidelberg Collaboratory for Image Processing, Visual Learning Lab,\\  Berliner Str. 43, 69120 Heidelberg, Germany}

\author[0000-0001-7906-3829]{Guido de Marchi}
\affiliation{European Space Research and Technology Centre, Keplerlaan 1, 2200 AG Noordwijk, Netherlands}

\author[0000-0002-3925-9365]{C.-H. Rosie Chen}
\affiliation{Max-Planck-Institut für Radioastronomie, Auf dem Hügel 69, D-53121 Bonn, Germany}

\author{Michele Cignoni}
\affiliation{Department of Physics - University of Pisa, Largo B. Pontecorvo, 3 Pisa, 56127, Italy }
\affiliation{INFN, Largo B. Pontecorvo 3, 56127, Pisa, Italy}
\affiliation{INAF-Osservatorio di Astrofisica e Scienza dello Spazio, Via Gobetti 93/3, 40129, Bologna, Italy}

\author{Andrew E.\ Dolphin}
\affiliation{Raytheon, 1151 E. Hermans Road, Tucson, AZ 85706, USA}
\affiliation{Steward Observatory, University of Arizona, 933 North Cherry Avenue, Tucson, AZ 85721, USA}
)

%%%%%%%%%%%%%%%%%%%%%%%%%%%%
%%%%%%%%% ABSTRACT %%%%%%%%%
%%%%%%%%%%%%%%%%%%%%%%%%%%%%

\begin{abstract}

In order to better understand the role of high-mass stellar feedback in regulating star formation in giant molecular clouds, we carried out a Hubble Space Telescope ({\sl HST}) Treasury Program {\sl Measuring Young Stars in Space and Time} ({\sl MYSST}) targeting the star-forming complex N44 in the Large Magellanic Cloud (LMC). Using the F555W and F814W broadband filters of both the ACS and WFC3/UVIS, we built a photometric catalog of 461,684 stars down to $m_\mathrm{F555W} \simeq 29$\,mag and $m_\mathrm{F814W} \simeq 28$\,mag, corresponding to the magnitude of an unreddened 1\,Myr pre-main-sequence star of $\approx0.09\,M_\sun$ at the LMC distance. 
In this first paper we describe the observing strategy of MYSST, the data reduction procedure, and present the photometric catalog. We identify multiple young stellar populations tracing the gaseous rim of N44's super bubble, together with various contaminants belonging to the LMC field population. We also determine the reddening properties from the slope of the elongated red clump feature by applying the machine learning algorithm {\sl RANSAC}, and we select a set of Upper Main Sequence (UMS) stars as primary probes to build an extinction map, deriving a relatively modest median extinction $A_{\mathrm{F555W}}\simeq0.77$\,mag. The same procedure applied to the red clump provides $A_{\mathrm{F555W}}\simeq 0.68$\,mag. \\\\

\end{abstract}

%%%%%%% KEYWORDS %%%%%%%
%\keywords{editorials, notices --- 
%miscellaneous --- catalogs --- surveys}

%%%%%%%%%%%%%%%%%%%%%%%%%%%%
%%%%%%%% MAIN TEXT %%%%%%%%%
%%%%%%%%%%%%%%%%%%%%%%%%%%%%
\section{Introduction}
The physical processes leading to star formation (SF) in the dynamically evolving multi-phase interstellar medium (ISM) are largely regulated by massive stars. In a star-forming region, the momentum and energy feedback from the few massive newborn stars is expected to terminate SF locally, trigger new SF remotely, and through gas expulsion modulate the gravitational potential and therefore stellar dynamics and cluster survival \cite[see e.g.\ the reviews by][]{MacLow04, Zinnecker07, McKee07}. Giant star-forming regions, aggregates of stellar nurseries spread over the scales of molecular clouds, are thus complex ecosystems where different stellar populations are born and interact with each other and their ambient ISM (for further discussions of the physical processed influencing ISM dynamics, see e.g.\ \citeauthor{Klessen16}~\citeyear{Klessen16} or \citeauthor{Girichidis20}~\citeyear{Girichidis20}). Recent studies of high-mass star-forming regions show significant sub-structure and hierarchical SF \citep{Bik2012, Gouliermis2014, Adamo2015, Sabbi2016, Cignoni2016, Gennaro2012, Gennaro2017, Nayak2016, Nayak2018, Sun2017, Getman2018, Dib2019, Grasha2019}, with star formation rates (SFR) and efficiencies (SFE) that vary within the same molecular cloud \citep{Hony2015}. Numerical modeling qualitatively reproduces this behavior \cite[e.g.][]{Bonnell97,Bonnell01, Klessen00, Offner09, Girichidis11, Girichidis12, Federrath12, Federrath13, Parker16,  Hennebelle18, Padoan20}, but our current understanding lacks the quantitative study of two critical measures needed to parameterize clustered SF: length-scale and time-scale. 
    
    %We propose to map the SF process in time and space, using the rich census of newly-born low-mass stars across a star-forming complex. In order to achieve its goal this analysis requires deep stellar coverage, well into the sub-solar mass regime, over a wide field-of-view, with two well-characterized Hubble filters, corresponding to the standard V and I.
    
Two main theories of SF on molecular cloud scales have been proposed, where the traditional approach postulates that stellar birth occurs in a slow quasi-static manner, with supporting mechanisms prolonging the cloud lifetime by many tens of dynamical times \cite[e.g.][]{Shu87, Krumholz07}, while the more modern dynamical theory of star formation acknowledges the complex morphological and kinematic structure of star-forming clouds and sees stellar birth as a highly dynamical, albeit inefficient, process. In this picture the formation of stars begins while the cloud is still forming and never reaches an equilibrium state before dispersing due to feedback \cite[e.g.][]{Hartmann01, BallesterosParedes07, Clark12, Chevance20}. These two scenarios can be observationally tested, both in terms of the morphological and kinematic properties of molecular clouds as well as in possible local variations of the stellar initial mass function (IMF), which on average exhibits remarkably uniform behavior \cite[][]{Kroupa02, Chabrier03, Bastian2010, Offner2014}. In quasi-static models, molecular clouds are globally gravitationally bound and well supported, allowing for slow SF, resulting in large age-spreads \citep[on the order of several dynamical timescales,][]{Shu1987,Tan2006}. %but little variation in the sub-solar initial mass function or SFE across the cloud {\bf [this needs a citation]}. 
In contrast, if the clouds are dynamically evolving and not necessarily globally gravitationally bound or very long lived, there will be a large variety of physical conditions including bound and unbound regions which will produce stars at both high and low efficiency, respectively. Furthermore, the sub-solar IMF can be significantly different as a function of the stellar clustering with a deficit of low-mass stars in the unbound, low SFE regions \citep{Bonnell2011}. %(Bonnell et al., 2011). 
    
To test these models one has to carry out a census of newly-born stars across a giant star-forming complex to identify and characterize each individual star-forming region over the whole field. In  particular, one would like to analyze the distribution of stellar ages, sub-solar IMF and SFE, and how these depend on the local gas properties.
While young stellar clusters are typically dominated by a handful of early-type stars already on the  main-sequence (MS), their main stellar population is largely composed by a multitude of intermediate- and low-mass stars in the pre–main-sequence (PMS), i.e.\ stars still in gravitational contraction toward the MS \citep[e.g.][]{Nota2006, Sabbi2008, Cignoni2009, Vallenari2010, Gouliermis2007, Gouliermis2011}. %(e.g. Nota et al., 2006; Sabbi et al., 2008; Cignoni et al., 2009; Vallenari et al., 2010; Gouliermis et al., 2007, 2011). 
Due to their relatively slow evolutionary time-scales (a 1\,$M_\sun$  star contracts to the main-sequence in $\sim 50$~Myr), they can be utilized as chronographs of the star formation history of the entire region. Therefore, while high-mass MS stars provide us with the signposts of ongoing star formation, it is the population of  intermediate-mass ($3\lesssim M/M_\sun\lesssim8$) 
Herbig Ae/Be and low-mass($M\lesssim3M_\sun$) T Tauri PMS stars that can provide us a direct measure of its youthfulness. In particular, 
by analyzing and comparing the different Hertzsprung-Russell diagrams one can chronologically sequence the recent star formation events, their duration, their mutual relations, and the possible differences between their stellar populations. %Therefore, the study of multiple populations of PMS stars, over a wide range of masses, enables a comprehensive analysis of the formation, evolution and interactions between young clusters, members of the same star-forming region. Such a study requires the unique Hubble sensitivity and resolution and cover different sub-solar populations over an area large enough to encompass the recent episodes of star formation, including their immediate surroundings.

Our neighboring galaxy, the Large Magellanic Cloud (LMC), provides the ideal environments for this study.  The LMC is a well-established laboratory to study SF because of its low metal abundance \citep[with a metallicity $Z\simeq 1/3$ Z$_\odot$, the LMC is a proxy of the early universe conditions at the cosmic noon of SF history, $z \sim 1.5$, e.g.][]{Madau1996}, low interstellar extinction \citep{Gordon2003}, and high SF activity. 
We have focused our attention on the LMC H II complex LH$\alpha$ 120-N44 \citep[][in short N44]{Henize1956}, with its rich ensemble of H II regions, bubbles, and young stellar clusters. 
The massive stars of the OB association LH 47 \citep{LuckeHodge1970}, located in the central super-bubble of N44, are the primary drivers of the expansion of the main bubble \citep{OeyMassey1995}. X-ray observations reveal $T\sim10^6$\,K gas heated by fast stellar winds and supernova explosions \citep{Jaskot2011}.  
The effects of stellar energy feedback, in particular along the western rim of the bubble where star formation may have been triggered by its expansion, are also evident through its H$\alpha$ and Spitzer images \citep{Chen2009, Carlson2012}. Herschel dust mass maps reveal the complex hierarchical ISM structure of N44 \citep{Hony2010}, and CO surveys show that star formation activity arises from one molecular cloud complex \citep{Fukui2001, Wong2011}, which can be analyzed for the process of hierarchical star formation. 
N44 has a total H$\alpha$ luminosity that places it between 30 Dor, an exceptional starburst event also in the LMC, and M42 in Orion, our closest example of ongoing massive star formation \citep[30 Dor : N44 : Orion = 20 : 1 : 0.04;][]{KennicuttHodge1986}. In conclusion, with multiple star-forming “hotspots” at different evolutionary stages, this complex provides the best paradigm of a “quiescently active” star-forming ecosystem. 

In this paper we present the first results from a new Hubble Space Telescope (HST) Treasury Program {\sl Measuring Young Stars in Space and Time} (MYSST, GO14689, P.I. D. Gouliermis). In Section \ref{sec:Observations} we summarize the observational parameters of the MYSST survey and the data processing strategy leading to the construction of the photometric catalog. In Section \ref{sec:StellarPopulations} we introduce the different stellar populations that can be isolated in the dataset. In Section \ref{sec:ExtinctionCorrection} we derive the optical extinction properties of N44 from the MYSST data by evaluating the slope of the reddened red clump feature in the colour magnitude diagram (CMD) and our method to assign a value of extinction to each source. Lastly, in Section \ref{sec:summary} we discuss and summarize our findings, concluding with an outlook on our future follow-up studies.

\section{Observations}
    \label{sec:Observations}
    \begin{figure*}
        \centering
        \includegraphics[width = 0.8 \linewidth]{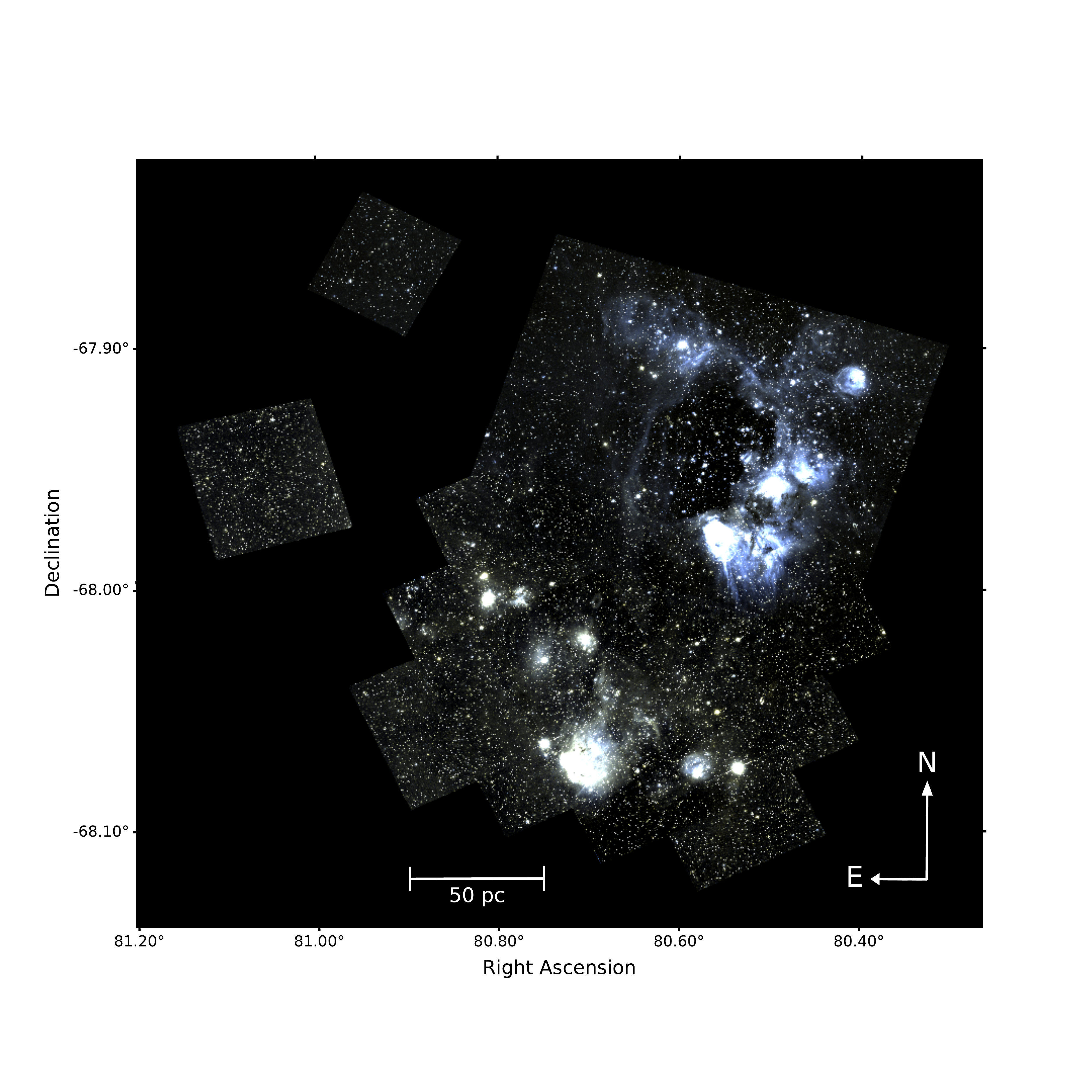}
        \caption{Two color composite image of N44 from the MYSST HST survey with the observations in F555W in blue and F814W in green. N44's characteristic super bubble can be seen in the North.}
        \label{fig:MYSST_color_composite}
    \end{figure*}
    
    \begin{figure}
        \centering
        \includegraphics[width = \linewidth]{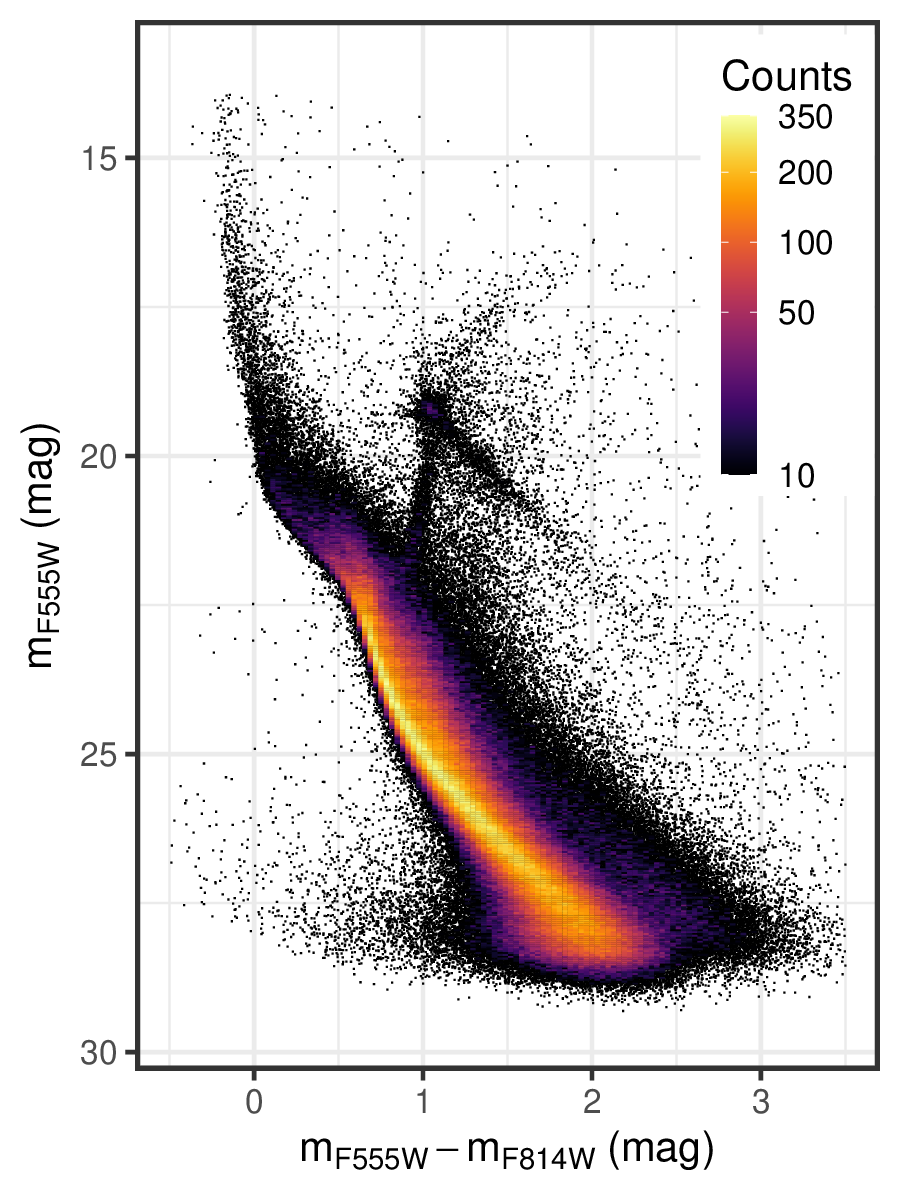}
        \caption{Optical CMD of the MYSST photometric catalog of N44. To highlight the structure of the CMD a 2D histogram with square bins of size 0.032\,mag is overlaid on the scatter plot where the number of stars per bin exceeds 10.}
        \label{fig:MYSST_CMD}
    \end{figure}   
    
    \begin{deluxetable*}{ccccc}
        \centering
        \tablecaption{MYSST Observations - Program 14689 \label{table_obs}}
        \tablehead{
        \colhead{Camera} & \colhead{Filter} & \colhead{Tot. EXPTIME} & \colhead{Short Exp.?} & \colhead{Visit Numbers}
        }
        \startdata
        \multirow{4}{*}{WFC3/UVIS} & \multirow{4}{*}{F555W} & 2715s & No & 01-12 \\
         & & 2532s & No & 25, 27 \\ 
         & & 2396s & Yes & 13-24 \\
         & & 2643s & Yes & 26 \\
         \hline
         \multirow{4}{*}{WFC3/UVIS} & \multirow{4}{*}{F814W} & 2532s & No & 01-12 \\
         & & 2808s & No & 25, 27 \\
         & & 2040s & Yes & 13-24 \\
         & & 2146s & Yes & 26 \\
        \hline
        \hline
        \multirow{4}{*}{ACS/WFC} & \multirow{4}{*}{F555W} & 2558s & No & 01-12 \\
         & & 2522s & No & 25, 27 \\ 
         & & 2361s & Yes & 13-24 \\
         & & 2517s & Yes & 26 \\
         \hline
        \multirow{4}{*}{ACS/WFC} & \multirow{4}{*}{F814W} & 2522s & No & 01-12 \\
         & & 2682s & No & 25, 27 \\ 
         & & 2030s & Yes & 13-24 \\
         & & 2020s & Yes & 26 \\
        \enddata
    \end{deluxetable*}

    Complementing the HST Treasury Programs on 30Dor \citep[GO-12939, P.I. E. Sabbi][]{Sabbi2013,Sabbi2016} and M42 \citep[GO-10246, P.I. M. Robberto,][and GO-12825, P.I. J. Shull (no refs yet)]{Robberto2013}, MYSST (GO-14689, PI: Gouliermis) is a deep, high spatial resolution HST survey of the star-forming complex N44 \citep{Henize1956} located in the Large Magellanic Cloud. It covers the large super bubble of N44 as well as the region south of it with a field of view (FoV) of $ 12.2 \times 14.7\,\mathrm{arcmin^2}$ that translates to about $180\,\mathrm{pc} \times 215\,\mathrm{pc}$ at the distance of the Large Magellanic cloud  assuming $(m-M)_0 = 18.55 \pm 0.05$ \citep{Panagia1991, DeMarchi2016}. The survey provides observations in two broadband filters, F555W and F814W, with the Advanced Camera for Surveys (ACS) and Wide Field Camera 3 (WFC3, UVIS channel) instruments of the HST.
    
    The N44 region was tiled in a grid pattern of 3 rows by 4 columns. Observations were taken in parallel, such that WFC3 covered the northern part of N44 and ACS the southern part, with a region of overlap in the middle. Table~\ref{table_obs} lists details of the observations we describe here. Each grid point was visited twice. Each visit consisted of two orbits, the first utilizing F555W and the second F814W with both cameras reaching down to 29 mag in F555W and 28 mag in F814W. In each orbit, four exposures were obtained using a subpixel box dither pattern. Two short (35s) exposures in each filter were obtained during the second visit to each grid point. Two additional fields to the east of the main mosaics were obtained by a single pointing with ACS and WFC3 observing in parallel. This pointing was visited three times with the same two-orbit, four-exposure setup. Two short (35s) exposures were obtained in each filter in the second visit. The HST two color composite image of N44 is shown in Figure \ref{fig:MYSST_color_composite}.

    \subsection{Data Processing}
    
    Bias, dark, flat-field, and charge transfer inefficiency corrected images, known as FLCs (\texttt{*\_flc.fits}), were downloaded from the Mikulski Archive for Space Telescopes (MAST)\footnote{\url{http://archive.stsci.edu/}}. These processing steps were performed by the standard calibration pipelines CALWF3\footnote{\url{https://wfc3tools.readthedocs.io/en/latest/wfc3tools/calwf3.html}} version 3.4.1 and CALACS\footnote{\url{https://www.stsci.edu/hst/instrumentation/acs/software-tools/calibration-tools}} version 9.2.0. The images were aligned to the Gaia reference frame \citep{gaiadr2} using \texttt{TweakReg}, part of the Drizzlepac software package\footnote{\url{https://www.stsci.edu/scientific-community/software/drizzlepac.html}}. The Gaia catalog was queried within the RA Dec bounds of the combined footprint of the FLC images, and the resulting sources were provided as a reference catalog to TweakReg to improve the absolute astrometry of our data. The coordinates of the Gaia sources span 81.1533 to 80.3041 degrees in RA and -68.1254 to -67.8355 degrees in Dec. The FLC images were aligned to better than 0.008'' (maximum root mean squared error).

    The long ($>$400s) FLC images were then combined using \texttt{AstroDrizzle} \citep{Hack2012} to create reference frames for each camera and filter for photometry. We used \texttt{resetbits}~$=4096$ to ignore existing cosmic ray flags, \texttt{skymethod}~$=$~\texttt{localmin} for sky subtraction, and \texttt{combine\_type}~$=$~\texttt{imedian} to avoid flagging saturated stellar cores as cosmic rays. The final drizzled images are sky subtracted and normalized by exposure time (units of e$^-$/s). The final pixel scales are native to each instrument, i.e.\ 0.05$\times$0.05~arcsec$^2$/pixel and 0.04$\times$0.04~arcsec$^2$/pixel for ACS and WFC3, respectively. The cosmic ray flagging performed by \texttt{AstroDrizzle} was propagated back to the data quality (DQ) extensions of the input FLC images.
    
    \subsection{PSF Photometry}\label{psf}
    
    \begin{deluxetable}{ccc}
        \tablecaption{Photometry Groups \label{table_groups}}
        \tablehead{
        \colhead{Group} & \colhead{ACS Visits} & \colhead{WFC3 Visits}}
        \startdata
        Strip 0 & -- & 05-12, 17-24 \\
        Strip 1 & -- & 01-08, 13-20 \\
        Strip 2 & 09-12, 21-24 & 01-04, 13-16 \\
        Strip 3 & 05-12, 17-24 & -- \\
        Strip 4 & 01-08, 13-20 & -- \\
        Field 0 & -- & 25-27 \\
        Field 1 & 25-27 & --  \\
        \enddata
    \end{deluxetable}
    
    \begin{table}[]
        \centering
        \caption{DOLPHOT Parameters for PSF Photometry}
        \begin{tabular}{ll}
        \hline
        \hline
        \multicolumn{2}{c}{DOLPHOT Parameters} \\
        \hline
        img\_rchi $=$ 2.0 & FSat $=$ 0.999 \\
        img\_raper $=$ 3 & PSFPhot $=$ 1\\
        img\_rsky $=$ 15 35 & FitSky $=$ 2 \\
        img\_rsky2 $=$ 4 10 & SkipSky $=$ 2\\
        img\_rpsf $=$ 15 & SkySig $=$ 2.25\\
        img\_apsky $=$ 20 35 & MaxIT $=$ 25 \\
        UseWCS $=$ 2 & NoiseMult $=$ 0.10\\
        Align $=$ 2 & SigPSF $=$ 3.0\\
        aligntol $=$ 4 & CombineChi $=$ 1\\
        alignstep $=$ 2 & DiagPlotType $=$ PS\\
        Rotate $=$ 1 & ApCor $=$ 1\\
        img\_shift $=$ 0 0 & Force1 $=$ 1\\
        img\_xform $=$ 1 0 0 & FlagMask $=$ 4\\
        SecondPass $=$ 5 & ACSuseCTE $=$ 0\\
        RCentroid $=$ 1 & WFC3useCTE $=$ 0\\
        SearchMode $=$ 1 & ACSpsfType $=$ 0\\
        SigFind $=$ 3.0 & WFC3IRpsfType $=$ 0\\
        SigFindMult $=$ 0.85 & WFC3UVISpsfType $=$ 0\\
        SigFinal $=$ 3.5 & InterpPSFlib $=$ 1\\
        PosStep $=$ 0.1 & PSFres $=$ 1\\
        dPosMax $=$ 2.5 & psfoff $=$ 0.0\\
        RCombine $=$ 1.415 \\
        \hline
        \end{tabular}
        \label{tab:dolphot_params}
    \end{table}
    
    Photometry was performed with DOLPHOT (version 2.0, downloaded on March 2, 2018\footnote{\url{http://americano.dolphinsim.com/dolphot/}}, see  \citeauthor{dolphin2000}~\citeyear{dolphin2000}), which is capable of running photometry on multiple images and cameras simultaneously. Because the full dataset is quite large, we split the visits into seven groups to maximize the photometric depth while minimizing the required computing resources and number of catalogs to merge. Table~\ref{table_groups} lists the visits contained in each group. The separated field visits were grouped by camera because they do not overlap the main mosaics, and were called Field 0 and Field 1. The main mosaic was divided into horizontal `Strips', called Strip 0 through 4. Strip 2 contains the region of overlap between the two cameras.
    
    The N44 star-forming region is fairly crowded, requiring PSF fitting photometry with DOLPHOT. TinyTim PSFs \citep{krist2011} included in the DOLPHOT download were used for both cameras. The simultaneous, iterative fitting and subtraction of stars by DOLPHOT refines the PSF model. The parameters used are given in Table~\ref{tab:dolphot_params}.

    Prior to running photometry, the SCI extensions of the drizzled reference frames and FLCs were masked according to the WHT and DQ extensions, respectively. The SATURATE header keyword was set to 71,000 e$^-$ for ACS FLCs and 55,000 e$^-$ for WFC3 FLCs because initial DOLPHOT runs were impacted by the presence of saturated pixels with values below the limits provided in the pipeline-processed files from MAST. Sky images were calculated for each FLC with \texttt{step}~$=-64$, $\sigma_{\mathrm{low}}=2.25$, and $\sigma_{\mathrm{high}}=2.0$. The F814W drizzled frame from the appropriate camera was used as the reference image for DOLPHOT alignment of the images in each Strip and Field; the WFC3 drizzled frame was used for Strip 2. Among all FLCs, the long exposures aligned to within 0.008'' and the short exposures to within 0.02''.

    Stellar sources, object types 1 and 2 (object type indicates a DOLPHOT internal source classification; values $\geq 3$ mark extended or single pixel sources), with $\mathrm{S/N}\geq5$ in both filters were selected from the full photometric catalogs to create intermediate so called ``\texttt{st}'' catalogs. In DOLPHOT object type 2 denotes 'star too faint for PSF determination'. This refers only to a position refinement procedure, as DOLPHOT uses different methods to measure position and fluxes of the detected sources. For type 1 the PSF is used to measure both flux and position, for type 2 the position from the initial finding stage is used instead.
    
    The ACS photometric system (VEGAMAG) and the WFC3 image coordinate system were chosen to be the survey standards. Well-measured stars in the region of overlap between the two cameras (Strip 2) were used to determine an empirical conversion from WFC3 to ACS magnitudes. In the expressions below $m_\mathrm{W555}$ ($m_\mathrm{W814}$) are the WFC3 F555W (F814W) magnitudes, and $m_\mathrm{A555}$ ($m_\mathrm{A814}$) are the ACS F555W (F814W) magnitudes. Given the WFC3 color,
    \small
    \begin{equation}
        C = m_\mathrm{W555} - m_\mathrm{W814},
    \end{equation}
    the ACS magnitudes are taken as
    \small
    \begin{align}
        m_\mathrm{A555} &= 
        \begin{cases}
          \!\begin{aligned}
            & m_\mathrm{W555} - 0.071 - 0.01(C - 1.45)  \\
    	    & \quad + 0.019(C - 1.45)^2
    	    \end{aligned}  & C < 1.45 \\[1ex]
        	m_\mathrm{W555} - 0.07 & C > 1.45 \\
        \end{cases} \\
        m_\mathrm{A814} &= m_\mathrm{W814} - 0.008.
    \end{align}
    \normalsize
    
    For the \texttt{st} catalogs of Strips 0 and 1 and Field 0, and the full catalog of Strip 2, the WFC3 photometry was converted to the ACS system. For Strip 2, the converted WFC3 photometry was combined with the ACS photometry with the same DOLPHOT technique used to combine multiple photometry blocks from individual FLCs (\texttt{CombineChi} $=1$). The \texttt{st} catalog criteria for object type and S/N were then applied to the full Strip 2 catalog. For Strips 3 and 4 and Field 1, the source coordinates in the \texttt{st} catalog were converted from the ACS reference image coordinate system to the WFC3 system using \texttt{astropy.wcs}, and corrected for a small residual offset ($-0.13$~pixels in $x$ and $-0.11$~pixels in $y$).

    The \texttt{st} catalogs containing ACS-system photometry and WFC3-system coordinates were merged by defining dividing lines between the Strips and Fields. Sources were retained from an individual \texttt{st} catalog if they fell in a given region defined by the dividing lines. Within $\pm$5~pixels of each dividing line, stars were matched between the two catalogs if the distance between their centers was $<$1~pixel and both magnitudes were within 0.25~mag. If a match was found, the coordinates were averaged and the photometry from the Strip appropriate for the average position was retained. Finally, sharpness (within $\pm$0.3) and crowding ($\leq$0.25 mag) criteria in both filters were applied to the combined \texttt{st} catalog to create the final PSF photometry catalog.

    \subsection{Aperture Photometry}
    
    Finalizing the PSF photometry catalog, we found that the brightest stars are saturated even in the short exposures. To recover their flux, we performed aperture photometry with DOLPHOT. Saturated stars exhibit a "bleed-out" effect into the neighboring pixels, but in the HST CCD detectors the photo-generated charges are conserved and using gain=2 the dynamic range of the CCDs is fully sampled by the Analog to Digital Converters. Thus, using a reasonably sized aperture, the total generated flux can be measured. To take advantage of the superior astrometry of HST and its well-understood PSF residuals, PSF photometry was first run using identical parameters as the previous runs, but only on the short exposures. For the aperture photometry, saturation and cosmic ray flags were ignored during the SCI extension masking step, and several DOLPHOT parameters were changed (listed Table~\ref{aper_params}). For each Field and Strip, the PSF photometry catalog from the short exposure run was specified in the UsePhot option for the aperture photometry run.
    
    \begin{deluxetable}{c}
    \tablecaption{DOLPHOT Parameters Updated for Aperture Photometry \label{aper_params}}
    \tablehead{
    \colhead{Updated DOLPHOT Parameters}}
    \startdata
    img\_raper $=$ 6 \\
    img\_rsky2 $=$ 7 12 \\
    SecondPass $=$ 1 \\
    RCombine $=$ 9 \\
    PSFPhot $=$ 0 \\
    FlagMask $=$ 0 \\
    \enddata
    \end{deluxetable}

    The same \texttt{st} selection criteria, magnitude and coordinate conversions, and catalog merging steps applied to the PSF photometry were applied to the aperture photometry. Less stringent sharpness (within $\pm$0.7) and crowding ($\leq$0.5 mag) criteria in both filters were applied to the combined catalog to create the final aperture photometry catalog.
    
    \subsection{Combined Photometry Catalog}

    Lastly, the final PSF and aperture photometry catalogs were merged together and assigned flags ($f_\mathrm{po}$) as described in the following steps. These steps were applied to stars satisfying $13.9\,\mathrm{mag} \leq \mathrm{F555W} \leq 18.3\,\mathrm{mag}$ and $12.9\,\mathrm{mag} \leq \mathrm{F814W} \leq 18.0\,\mathrm{mag}$. Brighter stars had saturated pixels extending beyond the 6-pixel aperture radius and were eliminated from the final, merged catalog, while fainter stars were better measured by PSF photometry (flag 4).
    
    First, magnitude offsets between PSF and aperture photometry were calculated from matching high S/N stars, and applied to the aperture photometry: $\Delta m_\mathrm{F555W} = 0.042$\,mag, $\Delta m_\mathrm{F814W} = 0.035$\,mag. Then, blends in the final aperture photometry catalog, i.e.\ multiple bright stars falling in the aperture, were identified. For each star, stars within 6 pixels in the \texttt{st} PSF catalog were found, and the potential aperture photometry contribution from neighbors was calculated from the difference between the combined brightness of all PSF photometry stars in the aperture and the brightest PSF magnitude. The star was determined to be a blend if the following criteria were satisfied for either filter:
    
    \begin{enumerate}
        \item the potential aperture photometry contribution from neighbors was more than 0.03 magnitudes,
        \item the aperture magnitude is brighter than the PSF magnitude of the brightest star and fainter than the brightest star minus twice the potential aperture photometry contribution from neighbors.
    \end{enumerate}
   \noindent
   This 0.03 mag threshold corresponds approximately to the apparent broadening of CMD features in the PSF-fitting photometry, i.e.~the error introduced in the PSF magnitudes by uncertainties in the precise PSF shape. The rationale of the other criteria is that contamination is given if the aperture photometry is consistent with the sum of the brightnesses of multiple objects within the 6 pixel radius.
    
    Lastly, for the final, merged catalog, aperture photometry was used for stars without matching sources in the final PSF catalog (flag 0) and for stars with a single match in the final PSF catalog (flag 1). PSF photometry was employed for blended stars in the aperture catalog (flag 2) and stars with no aperture detection (flag 3).
    
    In total the MYSST photometric catalog consists of 461,684 sources across the observed FoV of N44 as well as two smaller reference fields in the LMC. Figure \ref{fig:MYSST_CMD} shows the optical CMD of the survey. Due to saturation, the catalog does not include objects brighter than 14\,mag in F555W and 13\,mag in F814W. Consequently, some of the most massive O-type stars in the region are not part of the catalog. The faintest detected objects in the catalog reach down to about 29 mag in F555W and 28 mag in F814W. The noticeable broadening of the upper main-sequence (UMS) and the striking diagonal elongation of the red clump (RC) indicate that N44 is subject to a substantial amount of differential reddening.
    
    The MYSST photometric catalog and the four individual mosaics (F555W-ACS, F814W-ACS, F555W-WFC3 and F814W-WFC3) are available at the MAST archive as High Level Science Products via~\dataset[10.17909//t9-p5vg-ke50]{\doi{10.17909//t9-p5vg-ke50}}\footnote{\url{https://archive.stsci.edu/hlsp/mysst}}.
    The catalog (see Table \ref{tab:mysst_phot_cat} for an excerpt in the Appendix) lists for each source the survey internal ID (i.e. survey name combined with sexagesimal coordinates), pixel coordinates $X$ and $Y$, celestial coordinates \textit{R.A} and \textit{Decl.}, DOLPHOT object \textit{type}, and magnitude $m$, photometric error $\sigma$, and the (DOLPHOT) photometry flag $f$\footnote{These are bit flags, relevant for this catalog are 0, 'Star well recovered in the image'; 1, 'photometry aperture extends off chip'; and 2, 'too many bad or saturated pixels' (see also DOLPHOT manual, \url{http://americano.dolphinsim.com/dolphot/dolphot.pdf}).} for both the F555W and F814W filters. Also provided are the following DOLPHOT output parameters: the PSF fit quality parameter $\chi^2$, the signal to noise ratio \textit{SNR}, sharpness \textit{shrp}, roundness \textit{rnd} and crowding \textit{crwd}, once for each filter individually and once as a combined value over both. In the latter case also the major axis (if source is not round) \textit{mjaxdir} is listed.

    \subsection{Artificial Star Tests}
    
    Artificial star tests were run with DOLPHOT to measure completeness in the N44 region. The input artificial star list generated with \texttt{fakelist} largely matches the distribution within color-magnitude space of stars in the final PSF photometry catalog. The bounds in color-magnitude space were $14\,\mathrm{mag} < \mathrm{F555W} < 36\,\mathrm{mag}$ and $-2\,\mathrm{mag} < \mathrm{F555W} - \mathrm{F814W} < 6\,\mathrm{mag}$. To reach $\mathrm{F555W} \sim 36\,\mathrm{mag}$, the dimmest region of the real color-magnitude distribution is extended to dimmer magnitudes while retaining the same color distribution. Note that saturation effects are not modeled in DOLPHOT's artificial stars. Spatially, the artificial stars were randomly placed around the image by \texttt{fakelist}, then manually separated for each Strip and Field according to the dividing lines described in Section~\ref{psf}. ACS coordinates and magnitudes were converted to WFC3 coordinates and magnitudes for Strips 0, 1, and 2, and Field 0. DOLPHOT was run for each Strip and Field with the original PSF photometry parameters, but with the additional parameter FakeStars set to the appropriate input artificial star list. The output lists (\texttt{.fake} files) for Strips 0, 1, and 2 and Field 0 were converted back to ACS magnitudes.

    \subsection{Completeness}
    
    \begin{figure}
        \centering
        \includegraphics[width = \linewidth]{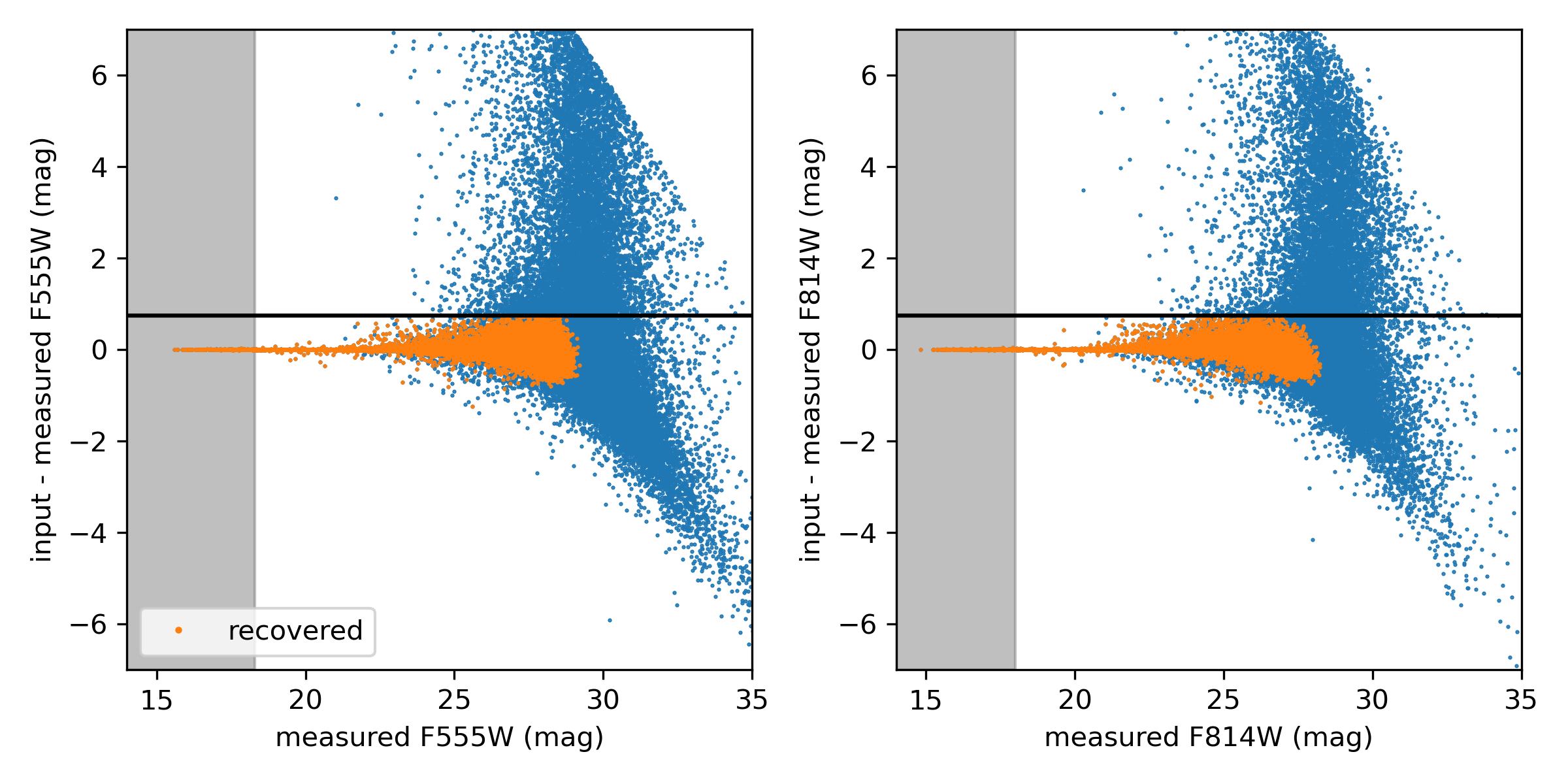}
        \caption{Difference between input and measured magnitudes as a function of measured magnitude of artificial stars for both F555W and F814W. The blue points are all artificial stars, and the orange points are those that satisfy our selection criteria for robust recovery, similar to the criteria used to accept stars in the data catalog. The horizontal line shows the $\mathrm{Input} - \mathrm{Measured} <0.75$~mag requirement to prevent co-location of recovered artificial stars with real stars of equal or brighter magnitude. The shaded regions indicate the magnitude ranges over which saturation prevents completeness measurements.}
        \label{fig:input_meas}
    \end{figure}  
    
    \begin{figure*}
        \centering
        \includegraphics[width = \linewidth]{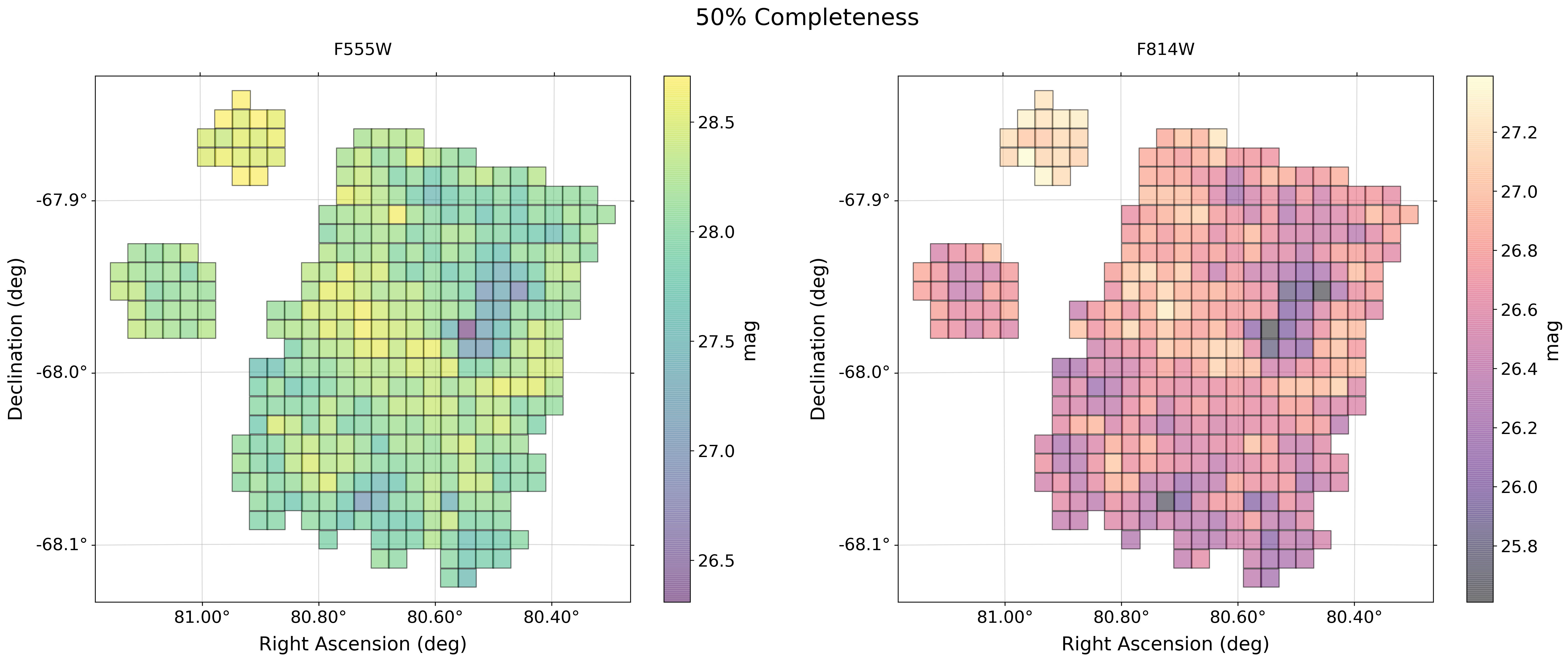}
        \caption{50\% completeness magnitude map for F555W (left) and F814W (right). Completeness magnitudes were measured in spatial bins of 1000$\times$1000~pixels.}
        \label{fig:comp50}
    \end{figure*}
    
    \begin{figure*}
        \centering
        \includegraphics[width = \linewidth]{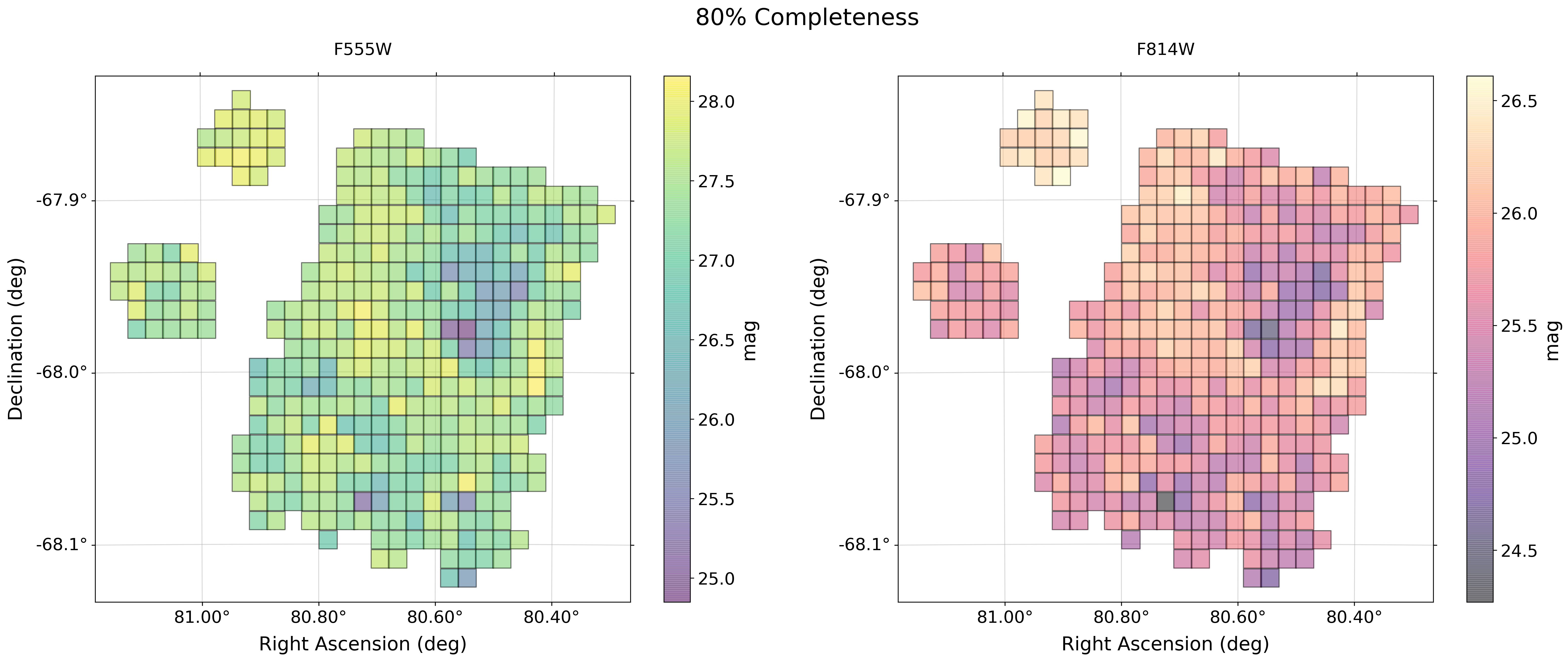}
        \caption{Same as Figure~\ref{fig:comp50} for 80\% completeness.}
        \label{fig:comp80}
    \end{figure*}
    
    To measure completeness, we first determine which input artificial stars were recovered by DOLPHOT. Stars were considered to be recovered if they met the same criteria in both filters for object type, S/N, sharpness, and crowding as the PSF photometry. A further recovery requirement on the artificial star photometry was that $\mathrm{Input} - \mathrm{Measured} < 0.75$~mag in both filters. This ensures that the artificial star was not co-located with a real star of equal or brighter magnitude. In Figure~\ref{fig:input_meas}, we plot the difference between input and measured magnitudes of the artificial stars as a function of measured magnitude. Recovered stars are highlighted in orange, and the $\mathrm{Input} - \mathrm{Measured} < 0.75$~mag requirement is shown as a horizontal line. The shaded regions indicate the magnitude ranges over which the PSF and aperture photometry catalogs were combined to capture saturated stars. The lower thresholds are $m_\mathrm{F555W} = 18.3~mag$ and $m_\mathrm{F814W} = 18.0~mag$. Because saturation effects are not properly modeled in DOLPHOT's artificial stars, we cannot study completeness effects in these regions.
    
    Completeness was measured in 1000$\times$1000~pixel spatial bins containing more than 500 stars across the main mosaics and offset fields. Within each spatial bin, we calculate the fraction of artificial stars recovered in 1 magnitude-wide bins in measured magnitude for each filter. We used linear interpolation of the recovery fraction over a finely-sampled magnitude range to find the 50\% and 80\% completeness magnitudes for each filter in each spatial bin. Figures~\ref{fig:comp50} and \ref{fig:comp80} show the 50\% and 80\% completeness maps of the N44 region, respectively. Spatial bins with brighter completeness limits correspond to regions of high stellar density, and in a few cases, saturated stars. The offset field and portion of the main mosaic covered by ACS/WFC show systematically brighter completeness limits. Over the entire N44 region, the average 50\% completeness is 28.1~mag for F555W and 26.7~mag for F814W, and the average 80\% completeness is 27.3~mag for F555W and 25.7~mag for F814W.
    
    Comparing the completeness in F555W to isochrones from the PARSEC stellar evolution models \citep{Bressan2012} of the appropriate metallicity for the LMC ($Z = 0.008$), we find that the 50\% and 80\% limits correspond to the brightness of unreddened $0.14\,M_\sun$ and $0.18\,M_\sun$ 1 Myr old pre-main-sequence stars at the distance of the LMC. The F555W detection limit of about 29 mag implies even a lowest mass limit of $0.09\,M_\sun$. Alternatively, for 10 Gyr\footnote{Note that these low-mass objects evolve so slowly that the mass limits derived from the 10~Gyr isochrone do not vary significantly from e.g.~either the 1~Gyr or 100~Myr isochrone, so are practically identical to the low-mass ZAMS value.} old, non-extinguished, low-mass main-sequence objects the 80\% and 50\% completeness limit, as well as the detection limit imply, mass thresholds of about $0.55\,M_\sun$, $0.5\,M_\sun$ and $0.4\,M_\sun$, respectively.

\section{Stellar Populations}
    \label{sec:StellarPopulations}

    \begin{figure*}
        \centering
        \includegraphics[width = \linewidth]{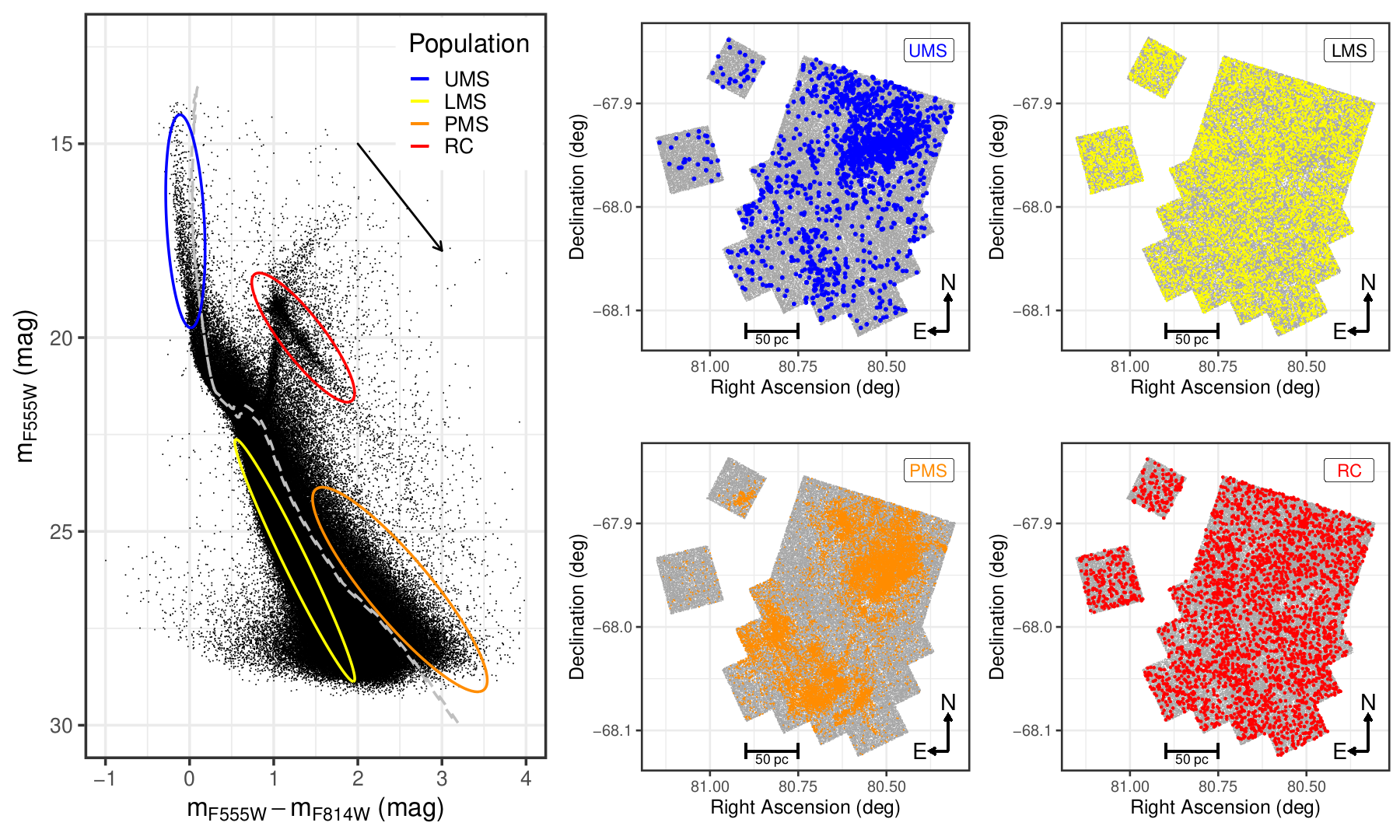}
        \caption{Optical CMD of the MYSST photometric catalog (left). Highlighted by the colored ellipses are the (rough) locations of different stellar populations found in the survey. For comparison, the grey dashed line indicates a 10 Myr PARSEC isochrone, corrected for the LMC distance modulus and the median UMS extinction derived in Section \ref{sec:UMS_ex}. The black arrow indicates the reddening vector as derived in Section \ref{sec:Ransac_ex}. Right: Spatial scatter plots of the roughly selected stellar populations shown in the CMD on the left in comparison to the total MYSST FoV. The population of LMS stars is sub-sampled to 10,000 examples for the diagram in the top right, as the LMS selection in the CMD contains more than 200,000 stars.}
        \label{fig:MYSST_Stellar pops}
    \end{figure*}
    In this section we shall briefly illustrate the stellar populations revealed by the MYSST survey. It is, however, not supposed to provide an exhaustive characterization. We further discuss the upper main-sequence and red clump stars in Section \ref{sec:ExtinctionCorrection} and dedicate a follow-up study \citep[][Paper II]{Ksoll2020} to the pre-main-sequence population of N44. 
    
    The rich CMD of the MYSST catalog (Figure \ref{fig:MYSST_Stellar pops}, left) immediately reveals that the survey captures a complex collection of different stellar populations. In the left panel of Figure \ref{fig:MYSST_Stellar pops} we highlight the loci of the most prominent stellar types (from an evolutionary standpoint). In addition, the right panel of this figure indicates the spatial distributions of these roughly selected stellar populations across the observed FoV. 
    
    The upper main-sequence (UMS) stars are young massive objects that trace the centers of star formation in large molecular clouds and rearrange their surrounding material through their powerful feedback in the form of winds and radiation \citep[i.e.][]{Elmegreen1977,Bisbas2011, Dale2013, Walch2013}. Several UMS stars show a color excess, likely being affected by significant reddening, although we cannot exclude that some of the most massive ones are rotating. In the latter case it is possible that they exhibit lower effective temperatures because of the lower internal pressure with respect to the non-rotating ones \citep[see e.g.][]{Meynet2000} or appear more luminous and hotter due to rotation induced internal mixing processes \citep[see e.g.][]{Brott2011}. In both scenarios rotation will induce an additional broadening of the UMS (and actually all stars with types earlier than F).
    
    Due to the previously mentioned saturation issues the MYSST catalog is likely missing the highest-mass O-type stars, such that our roughly indicated UMS population consists of late O, B and early A type objects. As the spatial distribution indicates, we find these sources predominantly in and around the massive super bubble of N44 in the northern half of the FoV. Additionally, there are a few compact clusterings south of the bubble indicating the presence of additional star-forming groupings. 
    
    The lower main-sequence (LMS) sources consist mostly of old, low-mass stars that belong to the LMC field population, lying either in the fore- or background of N44, contaminating the FoV of the MYSST survey. Most of these sources are too old to belong to the young star-forming centers of N44, but make it difficult to determine the population of still forming stars in the CMD. As extinction effects can dislocate their CMD position they may overlap with the CMD regions reserved to objects in the formation process. Note that differential distance effects are negligible for the LMC given its low scale height of 500 pc \citep{vanderMarel2001} in comparison to its distance of more than 50 kpc. Therefore, even a separation of e.g.~about 1 kpc between a star in front and one behind the LMC results in a magnitude difference smaller than 0.05 mag.
    
    If there have been previous star forming events in N44, before the currently observed star forming activity, it is also possible that older (e.g.~$> 50$ Myr) low-mass pre-main-sequence cluster stars, which are close to joining the MS, fall into the indicated LMS region in the CMD. Without additional measurements of e.g.~the $H_\alpha$ excess these still forming objects are notoriously difficult to distinguish from the old field LMS stars. The LMS sources are numerous, our rough selection containing already more than 200,000 objects, and as expected for field populations they are almost uniformly distributed (bar completeness and extinction modulation) as the sub-sampled (to 10,000 examples) spatial scatter plot reveals.
    
    The red clump (RC) population consists of old stars, already in their post-main-sequence evolution, that are in the process of burning helium in their cores. The luminosity of RC stars is almost insensitive to their age (at least for RC stars older than about 2 Gyr), thus they have been often used as standard candles to determine distances \citep[e.g.][]{Stanek1998, Girardi2001} and reddening \citep{Udalski1999a, Udalski1999b, Zaritsky2004, Haschke2011, DeMarchi2014, DeMarchi2016}. Like the LMS sources these, too, belong predominantly to the LMC field being fore- and background objects projected into the line of sight of N44. Under ideal observational circumstances, i.e.\ in the absence of reddening, these old post-main-sequence objects form an almost circular over-density in the CMD. Given that a fraction of the observed RC sources should be behind N44 we can infer from the notable elongation of the red clump feature in the observed CMD that N44 is host to a substantial amount of obscuring gas and dust that significantly reddens background sources and N44's constituents \citep{Dalcanton2015}. As the LMS objects, the RC stars are also mostly uniformly distributed across the MYSST FoV, confirmed by the spatial scatter plot in Figure \ref{fig:RC_ex_sel}. 
    
    Lastly, the pre-main-sequence (PMS) population is composed of very young objects that are still contracting under self-gravity, sometimes still accreting gas from their circumstellar disks and envelopes \citep{Manara2012}, and not yet dense and hot enough to ignite hydrogen burning in their cores. A comparison with the isochrone traced in the left panel of Figure \ref{fig:MYSST_Stellar pops} shows that our contour selects PMS stars likely younger than approximately 10 Myr. These actively forming stars make up the bulk of N44's young stellar clusters and characterize the star formation environment of this complex. As the survey title suggests, they are one of the primary targets of MYSST for recovering N44's star formation history. Our rough selection of the PMS population already highlights that N44 is host to a large number of young forming stars, that are predominantly located in and at the edges of the region's characteristic super bubble. Given that the bubble traces N44's gas reservoirs this arrangement is not surprising, but we also find additional compact clustered structures in the southern part of the survey. Our rough selection of PMS candidates only represents the most recent star formation activity in N44 and it is plausible that older (e.g.~$\sim20$-50 Myr) and less luminous PMS objects may be more spatially diffuse and confused with the LMS field population. In the immediate follow-up study to this introductory paper we quantify the young PMS population of N44 and characterize their clustering behavior \citep{Ksoll2020}.

\section{Extinction of N44}
    \label{sec:ExtinctionCorrection}
    In this section we study the extinction properties of N44 based on the photometric observations and construct an extinction map for the observed FoV. To achieve the latter, we estimate the extinction for a selection of UMS stars by projecting them along the direction of reddening onto their theoretical zero-age main-sequence (ZAMS) locus in the CMD. Afterwards, following a procedure we have outlined in \cite{Ksoll2018}, we assign a distance weighted average extinction of the 20 nearest UMS sources to the remaining stars in the catalog \citep[see also][]{DeMarchi2016}. For comparison we repeat the same procedure with RC extinction probes to further characterize the reddening profile of N44.  
    \subsection{Extinction Properties}
         One way to constrain the extinction properties from our photometric observations is utilizing the red clump feature in the CMD. Under perfect conditions, i.e.\ without extinction and photometric errors, the red clump is a well defined and easy to identify CMD feature. It is an over-density caused by core helium-burning post-main-sequence stars. Subject to differential extinction, however, the red clump appears elongated in the CMD tracing the reddening vector, i.e.\  
        \begin{equation}
            R_\mathrm{F555W-F814W}(\mathrm{F}) = \frac{A(\mathrm{F})}{\mathrm{E(F555W - F814W)}},
        \end{equation}
        where F indicates one of the two filters F555W or F814W.
        
        Therefore, the underlying extinction properties can be determined by measuring the slope of this CMD feature \citep[see e.g.][]{DeMarchi2014A, DeMarchi2016, DeMarchi2020}. To accurately retrieve this slope the constituents of the red clump need to be quantified first. Doing so by performing an arbitrary rectangular or elliptic selection around the red clump has the major drawback of inducing an uncontrolled  range of possible solutions. While this strategy may return a value that is not far off the true value, this approach is prone to non-objective selection effects. To circumvent this issue, e.g.\ \cite{DeMarchi2016} applied unsharp masking to constrain the red clump feature. In this paper we follow a different approach,  fitting the slope of the red clump while accounting for the presence of many outliers using a well established machine learning algorithm, RANSAC.
        \subsubsection{RANSAC}
            \label{sec:Ransac_ex}
            \begin{figure*}
                \centering
                \includegraphics[width = \linewidth]{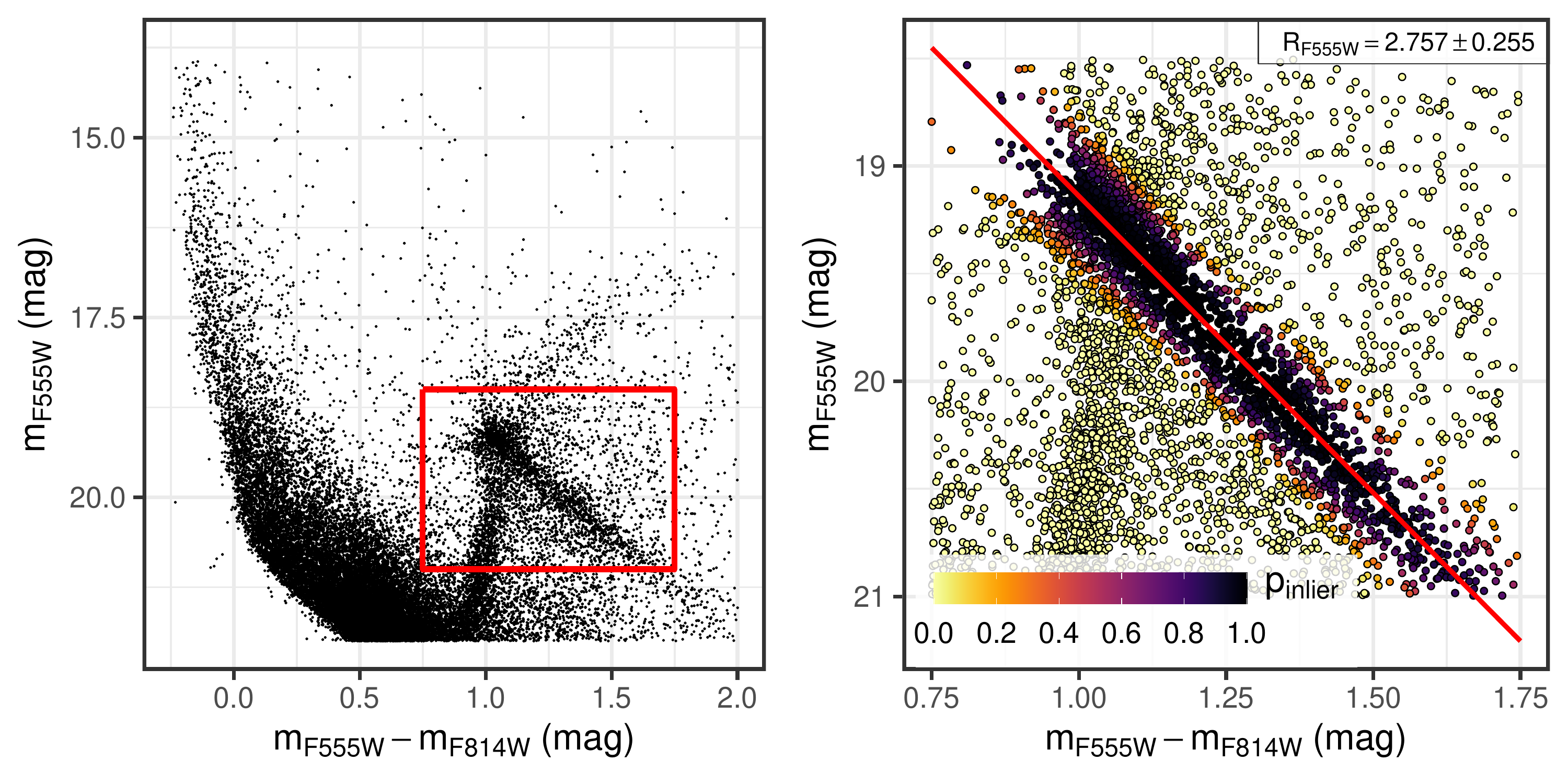}
                \caption{Zoom-in on the bright end of the optical CMD of the MYSST N44 data (left). Highlighted in red is the region used to derive the optical extinction law of N44 from the slope of the red clump elongation. Right: Zoom-in on the selected candidate red clump region from the left panel. The stars are color coded according to the fraction of times they were considered as a red clump inlier across the 5,000 RANSAC runs. The red line indicates the averaged resulting slope across all RANSAC fits. For the reddening vector in F555W we find a value of $R_\mathrm{F555W} = 2.8 \pm 0.3$.}
                \label{fig:RansacExtinctionLaw}
            \end{figure*}
            The RANSAC \citep[RANdom SAmple Consensus,][]{FischlerBollesRANSAC1981} algorithm can perform robust fits to datasets that suffer from outliers. The underlying assumption of this approach is that the data consists primarily of a set of inliers and a few outliers. Furthermore, this set of inliers is explained by a single model and the corresponding set of parameters, which we want to find by fitting the model to the data. To derive these parameters RANSAC selects a series of random subsets of the data and fits the model to each of these, deriving multiple sets of fit parameters. The idea is that, as the outliers are a minority in the data, most of these subsets consist only of inliers returning the same (or similar) fit parameters. In contrast the sets that contain random outliers will not agree on any given fit parameters. Therefore, by simply "counting the votes" for the parameters of all the random subsets, the underlying data generating model parameters are revealed. 
            
            In practice the algorithm first draws a random set of $n$ points, a minimum required to fit the desired model (e.g.\ 2 for a line), and performs a fit to that random subset. Then it determines the amount of remaining data that agrees with this fit, the inliers to this specific set of model parameters. To do so for every data point a distance to the fitted model is determined and compared to a preset acceptance threshold. If there are enough inliers to a model it is accepted as a good fit. This identified 'consensus' set is then employed to refine the fit by using all inlier points to re-estimate the model parameters. This procedure is repeated a number of times, $k$, to determine the best model as given by the fitting error. The number of random samples $k$ of size $n$ that have be drawn is chosen such that it has a low probability $p$ of containing only bad samples (in this study we use $p = 0.01$), and is given by the equation
            \begin{equation}
                k = \frac{\log(p)}{\log(1-w^n)},
                \label{eq:RANSAC_k}
            \end{equation}
            where $w$ is the fraction of inliers in the data. The parameter $w$ is of course often unknown, but by starting with a low estimate of $w$ the number of samples $k$ can be iteratively determined by updating the current guess of $w$ after every random sample with the actual determined fraction of inliers. This procedure stops when the number of samples that have been drawn exceeds the latest estimate of $k$ \citep{FischlerBollesRANSAC1981, Forsyth2003}. 
            
            In order to make use of the RANSAC algorithm to determine the extinction law of N44 we first need to make an initial selection in the CMD of the region where the elongated red clump is the dominant feature. The left panel in Figure \ref{fig:RansacExtinctionLaw} shows our selected region. We make this area large enough to be as agnostic as possible to our prior knowledge of where the red clump is located. In particular, the axes of the rectangle do not prescribe a slope. At the same time we make sure that the red clump is the major feature within the selection while we exclude a large portion of the red giant branch leading up to the red clump to facilitate the RANSAC inlier search. 
            
            Since our ultimate goal is to fit a line in the CMD, the RANSAC algorithm will draw samples with the minimum amount of necessary points, i.e.\ $n=2$ in our application. From a series of initial tests we determine that an acceptance threshold of 0.26 in the absolute error returns inlier selections and model fits which trace the elongated red clump as we would expect it. To minimize any influence from the random sampling on the final fit of the slope of the reddening vector, we repeat the RANSAC algorithm 5,000 times. Note that this is not the $k$ parameter: each of these 5,000 runs will draw $k$ random samples, with $k$ automatically determined according to Eq. \eqref{eq:RANSAC_k}. 
            
            As final value of the slope we take the average of the predicted slopes of these 5,000 RANSAC runs, and use their standard deviation as the associated uncertainty. The right panel in Figure \ref{fig:RansacExtinctionLaw} presents the result of this procedure for the slope of the reddening vector, $R_\mathrm{F555W-F814W}(\mathrm{F555W}) = 2.8 \pm 0.3$. Note that we color coded every star in this diagram according to the fraction of times across the 5,000 RANSAC runs that it was selected as an inlier to the final model.The plot shows that the RANSAC algorithm is able to accurately recognize the constituents of the elongated red clump feature. In fact, out of 5,000 runs the red-giant branch (RGB) leading up to the red clump is never identified as the dominant feature in our selection window. Using the same data selection and RANSAC procedure the corresponding slope in our other filter is $R_\mathrm{F555W-F814W}(\mathrm{F814W}) = 1.8 \pm 0.3$. For comparison the \cite{Cardelli1989} galactic reddening law, with $R_\mathrm{V} = 3.1$, returns $R_\mathrm{F555W-F814W}(\mathrm{F555W}) = 2.4$ and $R_\mathrm{F555W-F814W}(\mathrm{F814W}) = 1.4$, respectively, while the \cite{Fitzpatrick1999} model gives $R_\mathrm{F555W-F814W}(\mathrm{F555W}) = 2.2$ and $R_\mathrm{F555W-F814W}(\mathrm{F814W}) = 1.2$. So overall we find slightly larger $R$ values in N44, indicating a more "gray" reddening.
    
    \subsection{UMS Extinction}
        \label{sec:UMS_ex}
        \begin{figure*}
            \centering
            \includegraphics[width = \linewidth]{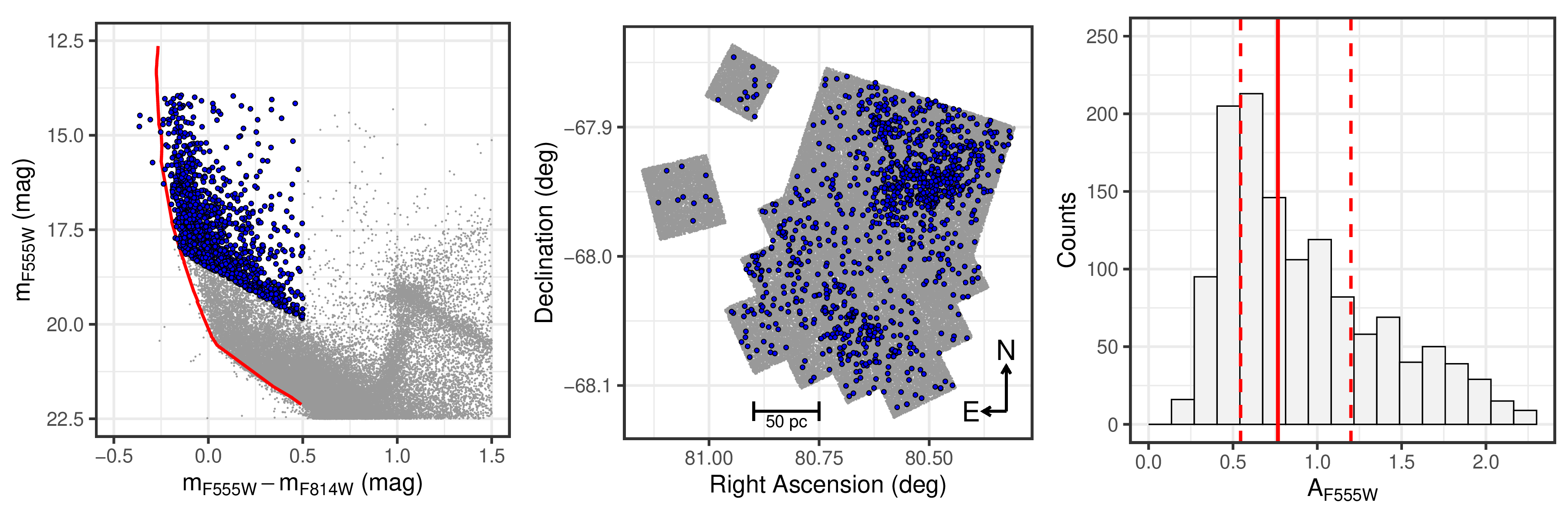}
            \caption{Zoom-in on the optical CMD of N44 (left). Highlighted in blue are UMS stars selected to estimate individual stellar extinctions. The red line indicates the position of the ZAMS, corrected for the LMC distance modulus and MW foreground extinction, that serves as the target position for the UMS extinction measurement. Spatial distribution of the MYSST photometric catalog of N44 (center). Highlighted in blue are the positions of our selected UMS extinction probes. Histogram of the extinction measurements (including the MW foreground) in F555W of the UMS probes (right). The solid red line indicates the median extinction of 0.77\,mag, while the dashed red lines mark the 25\% and 75\% quantiles.}
            \label{fig:UMS_Selection}
        \end{figure*}
        
        \begin{figure}
            \centering
            \includegraphics[width = \linewidth]{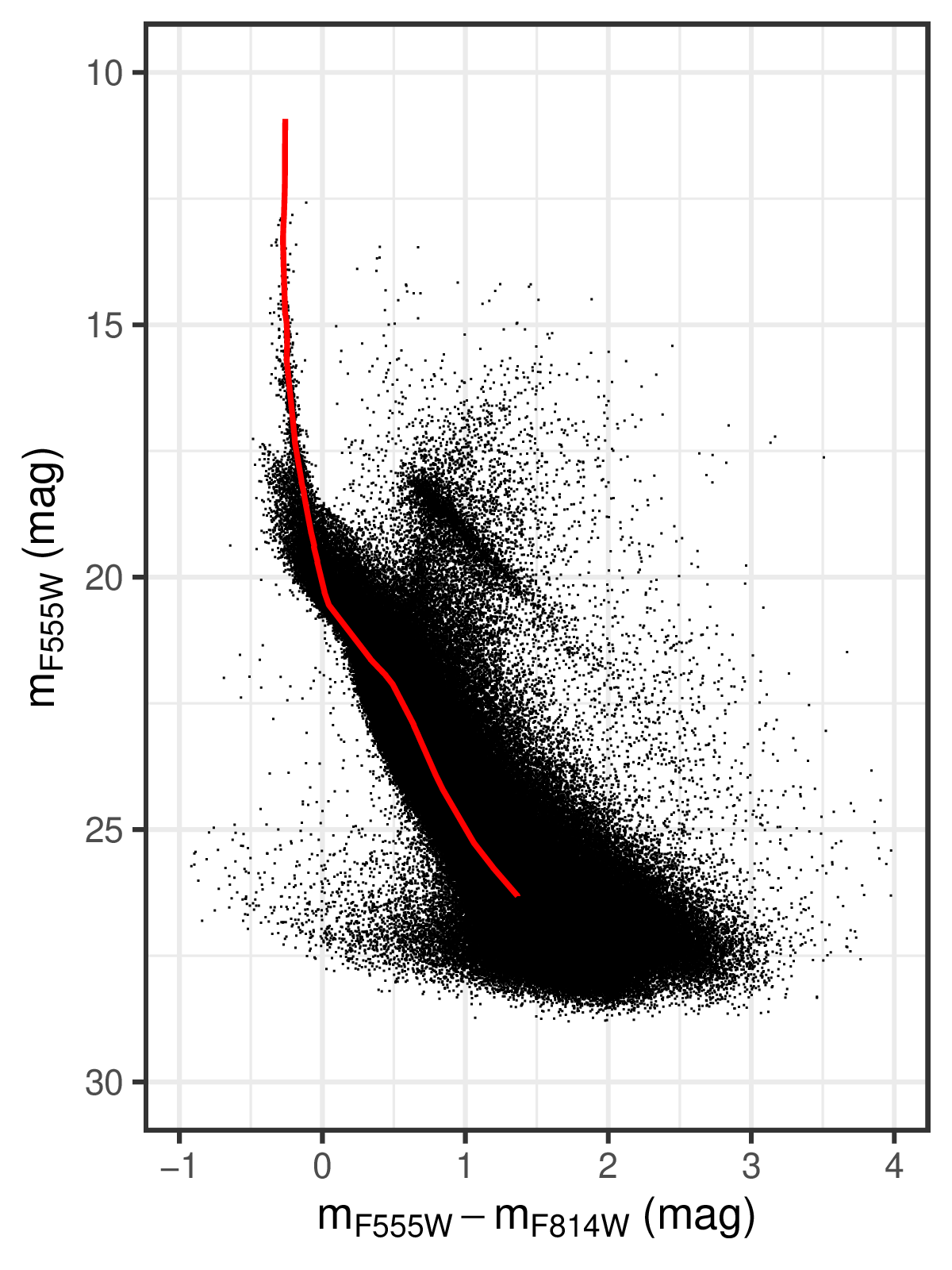}
            \caption{Optical CMD of the N44 data corrected for extinction. Each star is corrected by the distance weighted averaged extinction of their 20 nearest UMS neighbors. For visualization purposes the extinction of each UMS star is sampled within its measurement uncertainty. For reference the red line indicates the ZAMS, corrected for the LMC distance and MW foreground extinction, that is used to determine the extinction of the UMS probes. }
            \label{fig:MYSST_CMD_ExCorr}
        \end{figure}
        
        \begin{figure*}
            \centering
            \includegraphics[width = 0.75\linewidth]{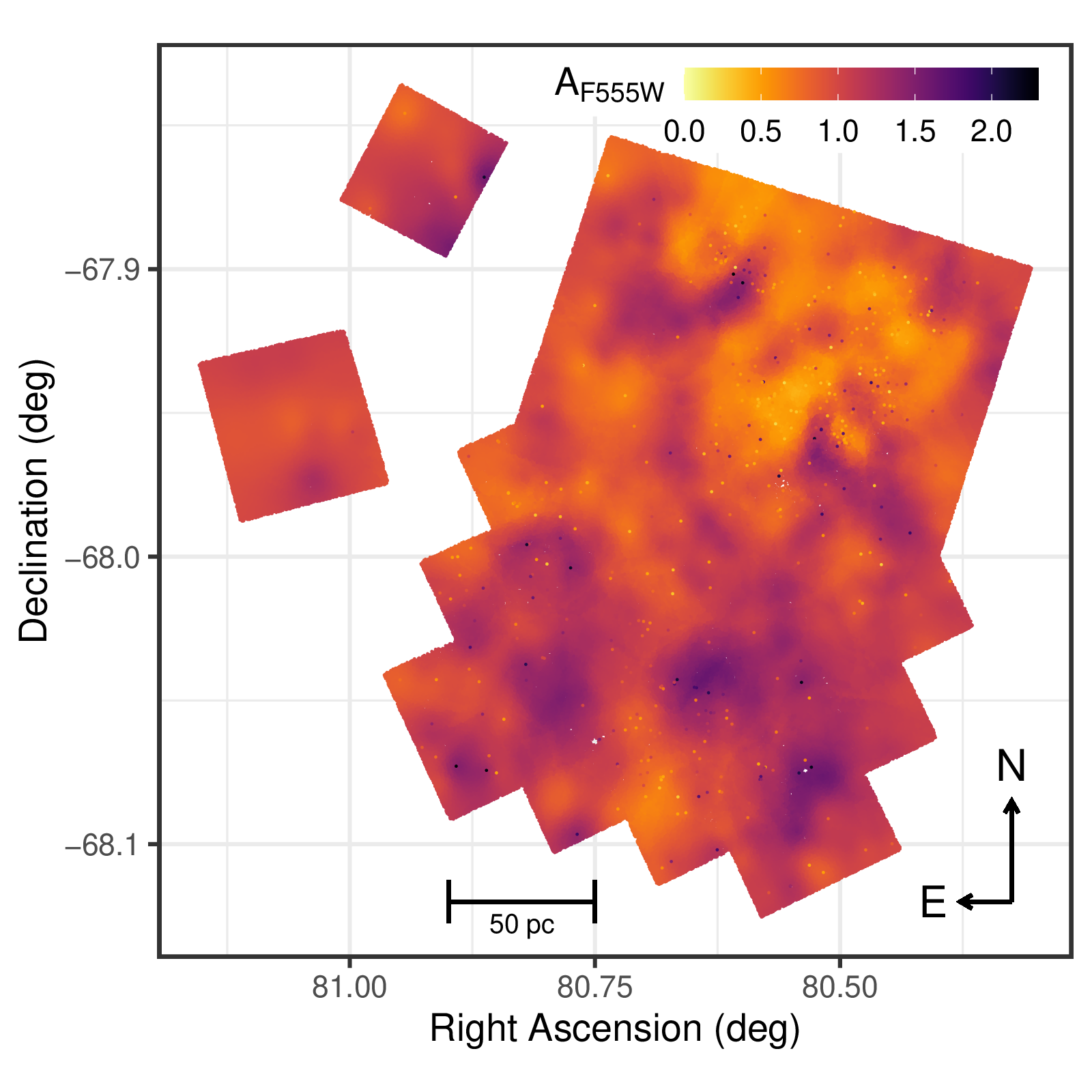}
            \caption{Spatial distribution diagram of the MYSST photometric catalog. Each star is color coded according to the assigned distance weighted average extinction of its 20 nearest UMS neighbors. The UMS extinction probes themselves appear as distinctly colored points in this diagram as they are not subject to the smoothing effect of our assignment procedure for the other stars. Additionally, any white spots are simply caused by a lack of sources, as this plot shows every individual star of the catalog.}
            \label{fig:ExtinctionMap}
        \end{figure*}
        With the reddening vector constrained we now measure extinction in F555W for UMS stars by re-projecting them along the reddening vector onto their theoretical optical CMD location, assuming that they should be on the ZAMS. Note that we derive the ZAMS locus from PARSEC isochrones \citep{Bressan2012} with a metallicity of $Z = 0.008$ for the LMC and correct the ZAMS position for Milky Way (MW) foreground extinction prior to the measurement. Here we adopt a value of $A_\mathrm{V}^\mathrm{mw} = 0.22$\,mag ($A_\mathrm{F555W}^\mathrm{mw}\approx0.223$\,mag) toward the LMC \citep{DeMarchi2014} and the \cite{Cardelli1989} Galactic reddening law. The measured value and the $0.223$\,mag MW foreground offset are then summed up to provide the total extinction. 
        
        For the extinction measurements we make a selection of the brightest UMS stars in the MYSST photometric catalog, accounting for the slope of the reddening vector and field contamination. This selection, depicted in the left panel of Figure \ref{fig:UMS_Selection}, consists of 1,291 stars, and represents a trade-off between retaining enough sources for good statistics and minimizing potential contamination from old field sources. Scaling the source density estimated in the reference fields to our CMD, we find that our selection criterion entails about 15\% field contamination, i.e.~specifically we expect $\mathbf{194\pm14}$ field stars in our sample. For more details on the UMS selection see Appendix \ref{app:UMS_selection}. As previously mentioned, due to saturation issues our selection of UMS stars is likely missing some of the most massive objects of N44. A comparison with the ZAMS, shifted along the reddening vector, puts our UMS sources in an approximate mass range between $\sim6$ and $\sim 30\,M_\sun$, indicating late O to early B-type stars. Note that there are 13 UMS stars in our selection with a CMD position that falls to the left of the target, foreground corrected,
        ZAMS locus. Given their close proximity to the ZAMS our procedure assumes that these sources have zero LMC extinction, so they are only subject to the MW foreground.  
        
        The middle panel of Figure \ref{fig:UMS_Selection} marks the positions of our UMS extinction probes in relation to the rest of the survey. As we can see, we find most of our UMS stars in and around the super bubble of N44 located in the northern part of the survey. With the fewer available probes in the middle and southern part of main FoV, as well as the very few UMS stars within the control fields, extinction estimates will be less precise there. 
        
        For the extinction measurement and assignment to the non-UMS stars we use a modified version of our approach presented in \cite{Ksoll2018}. In order to give a measurement uncertainty for the UMS extinctions, we now sample the slope of the reddening vector within its error and derive the mean and standard deviation of the so measured extinction values for each UMS star. Consequently, we now derive the error for the distance weighted average extinction assigned to the non-UMS stars by propagation of the uncertainty of the UMS measurements
        \begin{equation}
            \delta  A_\mathrm{F555W} = \sqrt{\sum_{i=1}^{20} \left(w_i \delta A_\mathrm{F555W}^\mathrm{UMS}\right)^2},
        \end{equation}
        with weights 
        \begin{equation}
            w_i = \frac{1}{d_i^2 + \epsilon^2} \frac{1}{\sum_{n = 1}^{20} (1/d_n^2 + \epsilon^2)}, 
        \end{equation}
        where $d_i$ is the Euclidean distance to the $i$th nearest UMS neighbor in pixels and $\epsilon$ is a smoothing factor \citep[for more details see][]{Ksoll2018}. 
        
        The distribution of the final F555W extinction measurements for the UMS stars is summarized in the right panel in Figure \ref{fig:UMS_Selection}. With an overall median extinction of $0.77_{-0.23}^{+0.42}$\,mag it appears that the UMS population of N44 is for the most part only moderately attenuated. This is consistent with the fact that a notable fraction of our UMS selection is located inside of the super bubble of N44, where feedback from the most massive stars has cleared out substantial amounts of gas. Still there are about two hundred UMS stars in our sample that exhibit more than 1.5\,mag of extinction up to a maximum of 2.29\,mag indicating the presence of regions that are subject to substantial reddening. 
        
        For the assignment to the non-UMS sources we follow our findings in \cite{Ksoll2018} and use the distance weighted average extinction of the 20 nearest UMS neighbors, employing a smoothing factor of $\epsilon = 500\,\mathrm{px}$ (20", $\sim 5$\,pc). Figure \ref{fig:MYSST_CMD_ExCorr} displays the CMD of the MYSST photometric catalog when corrected for the assigned extinction values in comparison to the ZAMS used to measure the extinction for the UMS stars. As we have elaborated in \cite{Ksoll2018}, this method of extinction estimation is not ideal for all types of stars, since we find some over- and underestimation for, e.g.~field MS and RC stars. The main issue here is that these objects are likely fore- and background objects of the young star-forming clusters that are the primary target of this survey. While recent studies \citep{Cignoni2015} provide strong evidence that young PMS objects tend to cluster around young massive UMS stars, so that this form of extinction estimate works well for young PMS objects \citep[see also][]{DeMarchi2016}, no such tendency for spatial co-location is given for the field contaminants. Therefore, the UMS extinction probes are less representative for these objects and the extinction estimate is less precise \citep{DeMarchi2011,DeMarchi2017}. 
        
        There is an additional caveat with our UMS extinction estimate to mention here. Our core assumption is that all UMS stars are on the ZAMS. Given the fast evolution of very massive stars, this might not necessarily be true and their actual position could be slightly different from the ZAMS, even if they are only a few Myr old. \cite{OeyMassey1995} find that the UMS stars interior to N44's super bubble are likely about 10 Myr old, while the ones located at the western rim of the bubble are 5 Myr old. Isochrones corresponding to these ages deviate to colors redder than the ZAMS in the high mass regime. 
        In these cases, dereddening to the ZAMS, i.e.~extending the vector beyond the correct isochrone, will overestimate both extinction and mass. Without additional information one cannot date the UMS stars more accurately across the entire MYSST FoV. However, due to the saturation limit of the MYSST survey, the number of stars where a significant difference between the ZAMS and the actual stellar age might occur is relatively small. Taking for instance a 10 Myr isochrone instead of the ZAMS, as appropriate for the bubble interior, only 276 stars out of our 1,291 UMS sources are massive enough to be affected. For the UMS stars located in the bubble and western rim we find a median absolute error of only $0.043_{-0.014}^{+0.019}\,\mathrm{mag}$ in extinction. Further details on this error estimate are provided in Appendix \ref{app:UMS_ex_error}. With this caveat in mind, our ZAMS assumption allows the derivation of a self-consistent (relative to the MYSST data) extinction estimate that in some cases may just provide an upper limit to the true value.
        
        Beside the age of the UMS there are other effects, e.g.~unresolved binarity or metallicity gradients, that can induce a broadening of the UMS in the CMD even in the absence of extinction. While often not considered for star clusters, there are findings that could support a potential metallicity gradient in N44. As previously mentioned, N44 has seen at least two known episodes of SF \citep{OeyMassey1995}. Additionally, a supernova remnant, SNR 0523-679 \citep{Chu1993}, which exhibits characteristics of a core collapse supernova \citep{Jaskot2011} is present within N44. \cite{OeyMassey1995} also estimate that up to four supernovae occurred in the region in the past. Lastly, \cite{Jaskot2011} find some evidence for metallicity enhancement in N44's super bubble. Thus, pollution of the formation environment of the younger population by one or multiple supernovae from the previous SF event is a possibility. We investigate the impact of these effects on our extinction estimation procedure in Appendix \ref{app:UMS_broadening}. 
        
        As a final test we also look into the potential extinction error induced by neglecting stellar rotation. As mentioned in Section \ref{sec:StellarPopulations}, rotation may induce additional color excess, in particular for very massive stars. To evaluate the effect we use the MIST \citep{Dotter2016MIST, Choi2016MIST, Paxton2011MIST, Paxton2013MIST, Paxton2015MIST} stellar evolution models (with $Z = 0.008$) to construct two ZAMS loci, one for the rotating ($v/v_{crit} = 0.4$) and one for the non-rotating case. Here we find that both ZAMS loci (and e.g. the 5 and 10 Myr isochrone, too) are practically identical, such that the expected extinction error caused by moderate stellar rotation is negligible for our approach. Using the SYCLIST\footnote{\url{https://www.unige.ch/sciences/astro/evolution/en/database/syclist/}} stellar evolution models \citep{Georgy2013} we also investigate the error induced in our approach for extremely fast rotating stars, i.e. $v/v_{crit} = 0.95$, in comparison to a non-rotating model. In this test we find a median absolute extinction error between the two models of $0.041_{-0.017}^{+0.022}\,\mathrm{mag}$. Consequently, the effect of rotation on our extinction estimate is minimal even for fast rotators. We have to note, however, that the SYCLIST models for $v/v_{crit} = 0.95$ are only available up to a stellar mass of $15~M_\sun$, such that the brightest 276 UMS stars in our selection could not be considered in this test. In addition we had to employ models with $Z = 0.006$ as the closest readily available metallicty to our adapted LMC value of $Z = 0.008$.
        
        Figure \ref{fig:ExtinctionMap} shows the extinction map we derive from our estimates by color coding each star according to its assigned value. As we can see the region of the super bubble is indeed subject to the least amount of reddening, while we find the most extinguished areas at the western edge of the bubble as well as predominantly in the southern part of the observed FoV. The rather prominent dark "filament", extending from the southern bubble edge to the south-west corner of the FoV, does not have a significant counterpart in the longer wavelength observations of N44 taken with Spitzer. Pointing to relatively low extinction values, this might suggest that this feature is actually an artifact of our ZAMS assumption. We verified that there are very few UMS stars in this "filament" that would experience a reduction in extinction measure if they were dereddened to a 10 Myr isochrone instead of the ZAMS (see Figure \ref{fig:UMS_ex_err} in Appendix \ref{app:UMS_ex_error}). Therefore, a significant systematic error could be justified only if these UMS stars are even older than 10 Myr.

        It should also be noted that the extinction of the two control fields is likely not particularly precise, due to the limitations of our approach listed above and the fact that these field stars are not likely to be co-located with the few selected UMS stars within these regions. 
    
    \subsection{RC extinction}
        \label{sec:RC_ex}
        \begin{figure*}
            \centering
            \includegraphics[width = \linewidth]{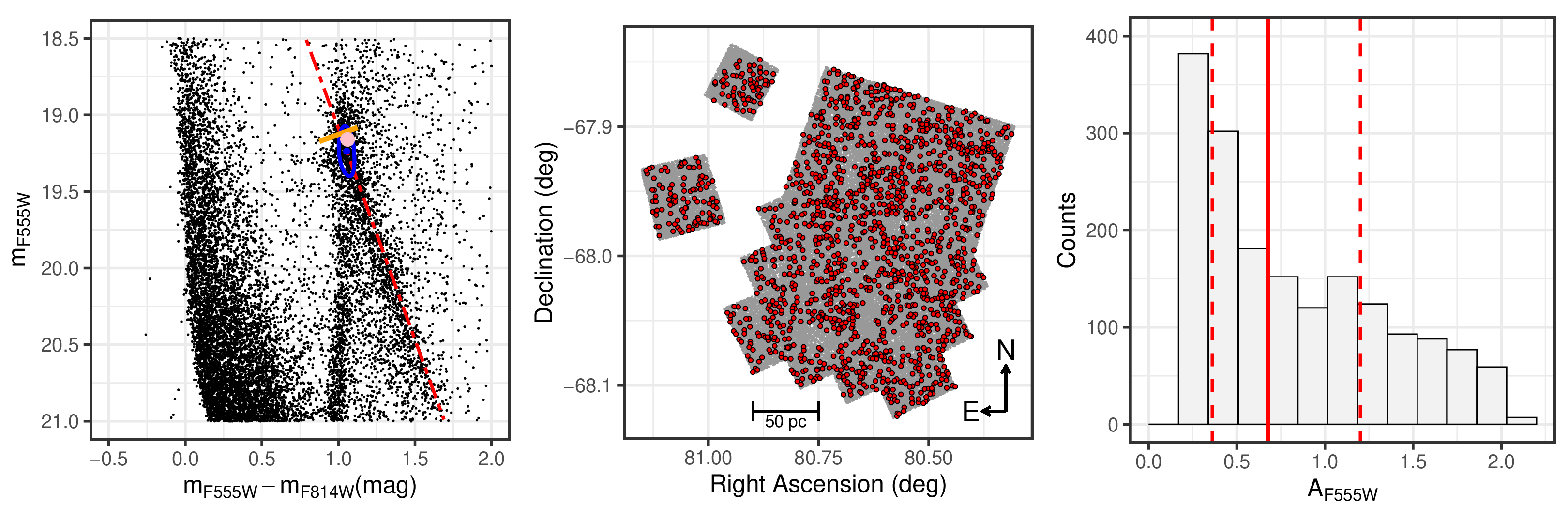}
            \caption{Zoom-in on the RC in the optical CMD of the MYSST catalog (left). The blue line represents a $4\sigma$ KDE density contour, marking the tip of the RC and indicating the nominal non-extinguished RC position in the CMD. The red dashed line marks the reddening vector quantified with the RANSAC approach. The solid orange line indicates the target true position, i.e.~the tip of the RC, used to determine the extinction of the red clump probes. It is perpendicular to the reddening vector and anchored to the intersection of reddening vector and the $4\sigma$ density contour. For comparison, the pink point indicates the theoretical RC position for the LMC field as determined by \cite{DeMarchi2014}, demonstrating the excellent agreement with our empirically determined position. Center: Spatial distribution of the RC constituents (red points) identified with the RANSAC procedure (i.e.~all sources with an inlier probability above 0.5). Right: Histogram of the measured extinction values in F555W of the RC probes. The solid red line indicates the median extinction of 0.68 mag, while the dashed lines mark the 25 and 75\% quantiles.}
            \label{fig:RC_ex_sel}
        \end{figure*}
        
        \begin{figure*}
            \centering
            \includegraphics[width = 0.45 \linewidth]{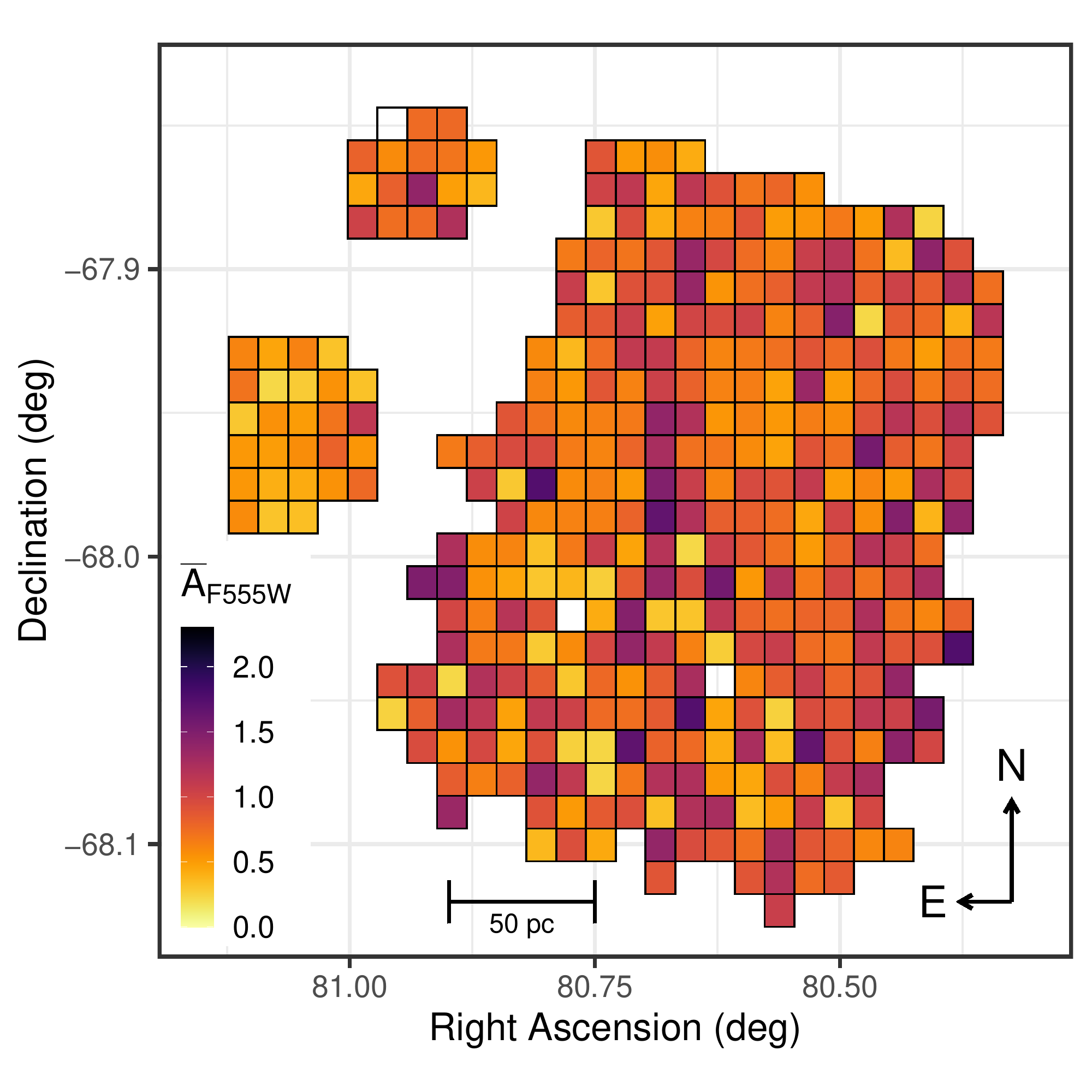}
            \includegraphics[width = 0.45 \linewidth]{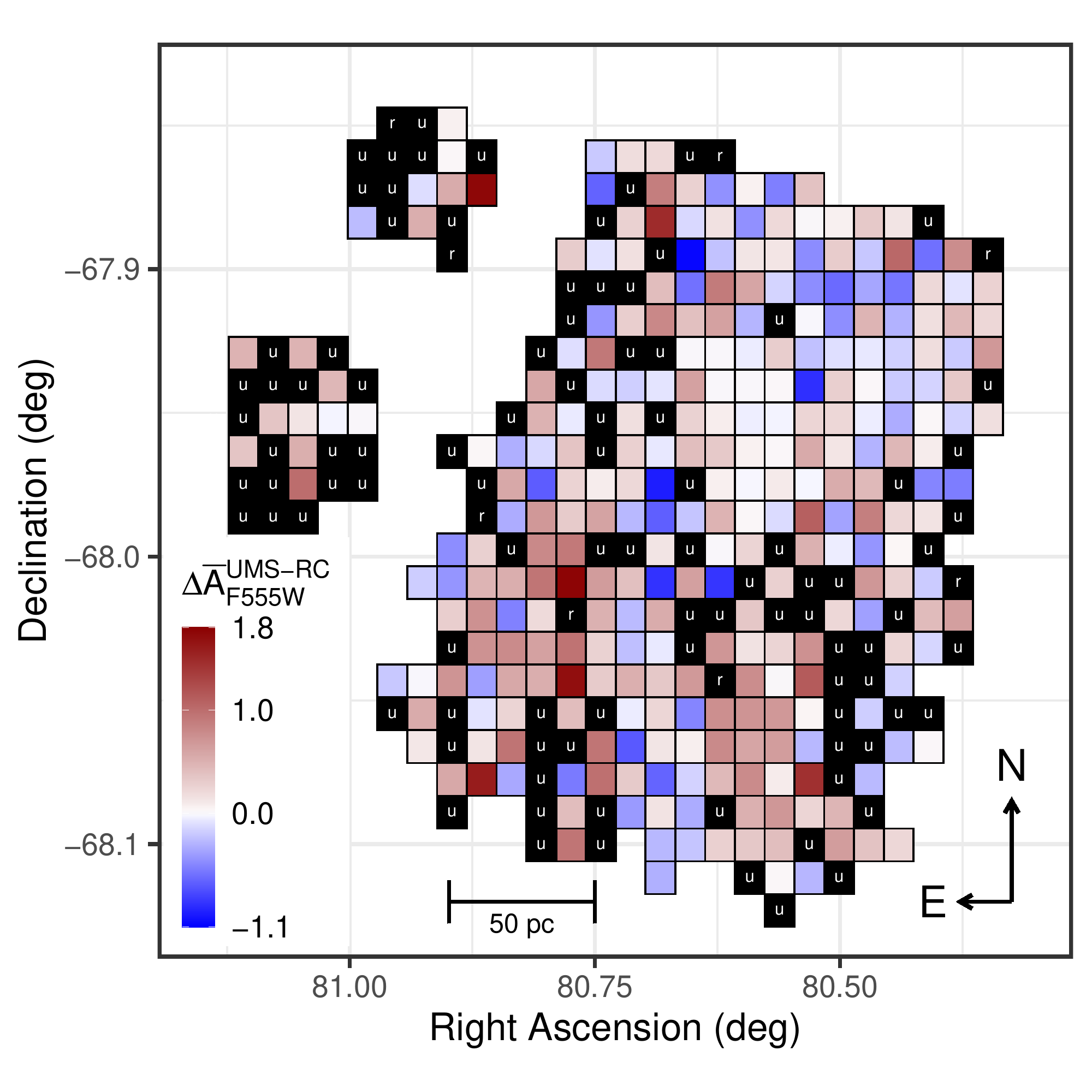}
            \caption{2D binned extinction map of the red clump sources (left). Each $41" \times 41"$ ($11\,\mathrm{pc} \times 11\, \mathrm{pc}$) bin is colored according to the average measured extinction of the red clump sources located inside. White tiles indicate bins in which no RC stars are found. Right: The same 2D bin diagram, but now each bin is colored according to the difference $\Delta \bar{A}_\mathrm{F555W}^\mathrm{UMS-RC} = \bar{A}_\mathrm{F555W}^\mathrm{UMS} - \bar{A}_\mathrm{F555W}^\mathrm{RC}$ of the mean extinction in each bin between the values derived from UMS and RC extinction probes. A positive value indicates a larger mean extinction derived from the UMS sources, while a negative one implies that the RC sources experience more extinction on average. The black tiles indicate bins where the difference $\Delta \bar{A}_\mathrm{F555W}^\mathrm{UMS-RC}$ cannot be computed because either no UMS ("u") or RC ("r") sources are present.}
            \label{fig:RC_EX_map_EX_diff}
        \end{figure*}
        
        To further characterize the extinction pattern of N44 we also determine extinction measures for the RC stars. As previously mentioned the non-extinguished RC is usually a prominent almost circular feature in the CMD, which is smeared out in the MYSST observation due to differential reddening. Still, the tip of this smeared out feature marks the nominal non-extinguished position of the RC. Consequently measuring extinction for the RC sources is also straight forward. To quantify the position of the RC tip we use a kernel density estimate (KDE) in the CMD space (using separate bandwidths for magnitude and color, each estimated by Silverman's rule) and identify the significant over-density in the vicinity of the tip. In particular we find a density contour at $4\sigma$ significance (above the mean density) which traces the non-extinguished end of the RC (blue contour in the left panel of Figure \ref{fig:RC_ex_sel}). We then determine the nominal RC position as a line perpendicular to the reddening vector anchored at the intersection point between the $4\sigma$ contour and the reddening vector going through its center. In practice, we vary the reddening slope within its uncertainty, determining a new target line for each sample slope, but keeping the same anchor point, in order to provide measurement errors for the RC probes. For comparison the left panel of Figure \ref{fig:RC_ex_sel} also indicates the theoretical position of the RC for the LMC field ($m_{F555W} = 19.16$, $m_{F814W} = 18.1$, corrected for distance modulus and foreground MW extinction) as determined by \cite{DeMarchi2014} for the HST filters used in this survey. Here we find an excellent agreement with our empirically determined nominal RC position. This match allows us to easily transform these relative RC extinction measurements to total extinction values by correcting for the Milky Way extinction contribution of $A_\mathrm{V}^\mathrm{mw} = 0.22$\,mag, translating to $A_\mathrm{F555W}^\mathrm{mw} \approx 0.223$\,mag, assumed for the theoretical RC position in \cite{DeMarchi2014}.
        
        From our RANSAC inlier determination, we select the bona fide RC constituents as those that reach an inlier fraction above 50\%, i.e.\ the stars that are chosen as RC inliers at least half of the time across all 5,000 RANSAC runs. This returns a sample of 1,737 RC stars, which appear to be almost uniformly distributed across the MYSST FoV (see Figure \ref{fig:RC_ex_sel}, middle panel, red points). As these objects are most likely fore- and background field sources and not part of the star-forming clusters in N44 this is to be expected. Note that there are 145 stars in our RC inlier selection which fall above our target RC position. Given their close proximity to the target our measurement approach assumes that these sources are only subject to the MW foreground extinction.
        
        The right panel of Figure \ref{fig:RC_ex_sel} shows the outcome of the extinction measurements for the RC probes. With a median total extinction of $0.68_{-0.32}^{+0.52}$\,mag RC stars are also overall only subject to moderate reddening. In comparison to the UMS sources the RC extinction distribution appears fairly similar being only about 0.1\,mag less extinguished on average. This slight difference could be due to the UMS sources being likely embedded within the star-forming centers of N44 whereas the mostly uniformly distributed RC field stars are not as obscured by N44's gas reservoirs. But with a maximum of 2.1\,mag extinction there are also a few hundred RC sources that are affected by more severe reddening. Note, however, that this maximum is certainly affected by our initially selected CMD region for the RANSAC procedure. It is possible that a few more heavily extinguished RC objects were excluded by this selection, so that this upper extinction limit should not be treated as an absolute maximum.
        
        To make a direct spatial comparison to the UMS extinction map we cannot follow the same procedure as for the UMS stars in Section \ref{sec:UMS_ex}. Assigning distance weighted average values of the nearest RC neighbors to the other stars has little meaning, since these RC field stars are very unlikely to have a spatial correlation to the N44 constituents beyond projection effects. Being predominantly part of the LMC field population the RC sources are, however, more likely to provide a representative extinction measure for other field sources, such as the many LMS stars captured in the MYSST FoV. To perform a spatial comparison we instead use a 2D binning approach, computing the average measured extinction of the RC sources in square $41" \times 41"$ ($11\,\mathrm{pc} \times 11\, \mathrm{pc}$) spatial bins. The resulting low resolution RC extinction map is shown in the left panel in Figure \ref{fig:RC_EX_map_EX_diff}. The most extincted RC stars are located towards the eastern and northern edge of the super bubble as well as in more compact regions south of the bubble. The right panel of Figure \ref{fig:RC_EX_map_EX_diff} provides a direct comparison between the 2D binned RC mean extinction map and a corresponding map derived from the UMS sources, showing the difference $\Delta \bar{A}_\mathrm{F555W}^\mathrm{UMS-RC} = \bar{A}_\mathrm{F555W}^\mathrm{UMS} - \bar{A}_\mathrm{F555W}^\mathrm{RC}$ in each bin. Positive values in this map indicate where the mean extinction inferred from the UMS stars is larger, while negative values imply the opposite. As we can see, the UMS extinction exceeds that of the RC mostly in the southern half of the FoV. Notable regions where the mean RC extinction is larger are located at the eastern and northern edge of the bubble as well as in several compact patches in the south of the FoV. These RC stars are likely located behind the star-forming gas of N44. On the other hand, in the regions where the mean UMS extinction dominates it is possible that background RC sources are simply not detected due to the obscuring gas that the UMS sources are embedded in. Interestingly, we find that the RC and UMS extinction agrees very well inside N44's bubble. This provides further confirmation that the feedback of the massive stars interior to the bubble has cleared out almost all of the gas, such that barely any local obscuration is left. It is necessary to note though that the difference between UMS and RC extinction might be affected by stochastic variability in the number of RC sources that are in the foreground and background in a given 2D bin in the right panel of Figure \ref{fig:RC_EX_map_EX_diff}.
        
\section{Summary}
    \label{sec:summary}
        
        In this paper we introduce the new HST Treasury Program 'Measuring Young Stars in Space and Time' (MYSST) which captures the active star-forming complex N44 with its rich collection of H II regions, young stellar clusters and bubbles, located in the Large Magellanic Cloud. We present the observing strategy of MYSST, describe our data reduction procedure and construct the photometric catalog of the survey. In addition we highlight our first scientific results, briefly discussing the stellar populations found across N44 and determining the extinction properties of the region. On top of that, we infer extinction maps for N44 from reddening measurements of UMS and RC stars. 
        
        The MYSST survey observed N44 in the optical wavelength regime using the F555W and F814W broadband filters of both the ACS and WFC3 imagers on board the HST. Combining PSF fitting and aperture photometry, the latter being needed to recover saturated bright sources, we compile a photometric catalog that comprises 461,684 stars across the MYSST FoV, going down as deep as 29\,mag in F555W and 28\,mag in F814W, probing even the lowest mass stellar population of N44 (e.g.~down to $0.09\,M_\sun$ for an unreddened 1 Myr pre-main-sequence star). Due to saturation effects the catalog does not contain sources brighter than 14\,mag in F555W and 13\,mag in F814W, likely missing the most massive O type stars of the region. Due to stellar crowding, background and saturation the completeness of the catalog varies across the FoV, but reaches an excellent average of 26.7 (28.1)\,mag in F555W and 25.7 (26.7)\,mag in F814W at the 80\% (50\%) level.
        
        The rich photometric catalog reveals many different stellar populations spread across the MYSST FoV. We identify numerous old lower main-sequence and red giant branch sources that are almost uniformly distributed across N44, likely fore- and background contaminants belonging to the LMC field population. We also find young high-mass upper main-sequence and lower mass pre-main-sequence stars within the survey, which exhibit clustered spatial distributions, tracing e.g.\ the gaseous rim of N44's characteristic super bubble. These young stars mark N44's numerous active star-forming centers.
        
        To constrain the reddening properties of N44 we measure the slope of the red clump feature that appears elongated in the CMD due to differential extinction. Here we present a new approach to jointly establish the constituents of this elongated red clump feature and perform the fit of the reddening vector by applying the well established learning algorithm RANSAC \citep{FischlerBollesRANSAC1981}. This algorithm is a very robust tool to fit models to data in the presence of outliers. RANSAC is an iterative process, where in each iteration first a minimal subset of the data, large enough to fit the given model, is drawn randomly from the total data. Then the model fit is performed on that subset and finally the amount of data points within the complete data set are determined that are inliers to the fitted model. These steps are then repeated until an optimal model is found at which point a final fit to the inliers of this model is performed for further refinement.
        
        Selecting a window within the optical CMD of the MYSST data in which the red clump is the predominant feature, we apply the RANSAC algorithm, repeating it 5,000 times to negate all effects of the random seed, to determine the red clump constituents and fit a line to the elongated red clump to derive the slope of the reddening vector. With this approach we find the total to selective extinction ratios $R_\mathrm{(F555W-F814W)}(\mathrm{F555W}) = 2.8 \pm 0.3$ and $R_\mathrm{(F555W-F814W)}(\mathrm{F814W}) = 1.8 \pm 0.3$ for the slope of the reddening vector in N44. These results are notably larger than the values for the standard galactic extinction law with $R_\mathrm{V} = 3.1$, returning 2.4 and 1.4 \citep{Cardelli1989} or 2.2 and 1.2 \citep{Fitzpatrick1999}, respectively. 
        
        With the reddening vector constrained we select a set of UMS stars as probes and measure their extinction by re-projecting their position in the CMD back to their theoretical location, assuming that they should be on the zero-age main-sequence. Afterwards we assign each non-UMS star in the MYSST photometric catalog a distance weighted average extinction of their 20 nearest UMS neighbors. This procedure has been found to return reasonable extinction estimates for the constituents of young star-forming clusters \citep{DeMarchi2016}, such as the ones we are aiming to find here, but suffers from occasional extinction over- or underestimation for field constituents. Additionally, this approach might overestimate extinction for older ($>10$\,Myr) very massive UMS sources, as their true position might slightly deviate from the ZAMS, such that at worst the estimate provides only an upper limit to the true extinction. For a subset of UMS stars with known ages inside and at the rim of N44's bubble we find, however, only a median absolute error of $0.043_{-0.014}^{+0.019}\,\mathrm{mag}$ with our approach compared to using the correct ages. We make the ZAMS assumption since the UMS ages are not easily recovered across the entire FoV and because it allows a MYSST self-consistent extinction estimate that entails the same systematic error everywhere. Note that we plan to provide more precise extinction measures in a follow-up study that explores synergies with other observations of N44 (e.g.~Gaia).
        
        Following the assignment, we present an extinction map for N44 based on the measured UMS extinction. With a median extinction of $0.77_{-0.23}^{+0.42}$ mag in F555W it appears that the UMS population of N44 is overall only moderately extincted. With about 200 UMS probes, though, exceeding $1.5$ mag up to a maximum of $2.29$ mag in extinction there is still a notable amount of regions subject to more severe reddening. Our extinction map confirms that the reddening of N44 is patchy and highly differential across the MYSST FoV.
        
        For comparison we also compile a 2D binned average extinction map derived from measurements of red clump stars. %The nearest neighbor approach used for the UMS map is not applicable here, as these field RC sources have no spatial correlation with the cluster constituents besides projection effects. 
        Showing an overall median extinction of $0.68_{-0.32}^{+0.52}$\,mag the red clump stars across the MYSST FoV are similarly reddened as the UMS population. There are a few hundred of our total $\sim$1,700 RC extinction probes that also exhibit more severe reddening up to a maximum of 2.1\,mag. This is, however, not an absolute maximum as our RANSAC approach may have excluded a few heavily extinguished RC stars. A direct spatial comparison between a UMS and RC 2D binned mean extinction map reveals that the UMS sources tend to be more reddened across most of the southern half of the FoV. Notable areas where RC extinction exceeds the UMS values are the eastern and northern edge of the N44 bubble, as well as a few compact patches south of the bubble.  
        
        In conclusion, the MYSST survey provides an extraordinary view of extragalactic star formation across an entire giant star-forming complex that highlights the complex interplay between high-mass stellar feedback and star-forming events. With its high resolution and deep photometry it provides the opportunity to study length and timescales of the star formation process on the scale of a giant molecular cloud. In our subsequent study in Paper II we begin to quantify the star formation history of N44 by identifying its rich PMS stellar content and analyze the complex clustering behaviour of the young still forming PMS stars across N44. 
        
\section{Acknowledgements}
    We would like to thank the anonymous referee for their thorough and timely review of our manuscript, as well as the detailed constructive feedback provided, allowing us to deliver a more complete and concise version of this study.\\
    The authors would like to thank Varun Bajaj for help with aligning and drizzling the data. \\
    VFK was funded by the Heidelberg Graduate School of Mathematical and Computational Methods for the Sciences (HGS MathComp), founded by DFG grant GSC 220 in the German Universities Excellence Initiative. VFK also acknowledges support from the International Max Planck Research School for Astronomy and Cosmic Physics at the University of Heidelberg (IMPRS-HD).\\
    RSK acknowledges financial support from the German Research Foundation (DFG) via the collaborative research center (SFB 881, Project-ID 138713538) {\em The Milky Way System} (subprojects A1, B1, B2, and B8). He also thanks for funding from the Heidelberg Cluster of Excellence {\em STRUCTURES} in the framework of Germany’s Excellence Strategy (grant EXC-2181/1, Project-ID 390900948) and for funding from the European Research Council via the ERC Synergy Grant {\em ECOGAL} (grant 855130) and the ERC Advanced Grant {\em STARLIGHT} (grant 339177). \\
    The project ``MYSST: "Mapping Young Stars in Space and Time'' is supported by the German Ministry for Education and Research (BMBF) through grant 50OR1801. \\
    Based on observations with the NASA/ESA Hubble Space Telescope obtained from the Mikulski Archive for Space Telescopes at the Space Telescope Science Institute, which is operated by the Association of Universities for Research in Astronomy, Incorporated, under NASA contract NAS5-26555. Support for program number GO-14689 was provided through a grant from the STScI under NASA contract NAS5-26555.

    \software{Astrodrizzle \citep{Hack2012}, 
    DOLPHOT \citep[v2.0][]{dolphin2000}, 
    PARSEC \citep{Bressan2012},
    MIST \citep{Dotter2016MIST, Choi2016MIST, Paxton2011MIST, Paxton2013MIST, Paxton2015MIST}, 
    SYCLIST \citep{Georgy2013}
    }

%%%%%%%%%%%%%%%%%%%%%%%%%%%%
%%%%%%%% REFERENCES %%%%%%%%
%%%%%%%%%%%%%%%%%%%%%%%%%%%%    
    
\bibliography{paper1.bib}{}
\bibliographystyle{aasjournal}

%%%%%%%%%%%%%%%%%%%%%
%%%%% APPENDIX %%%%%%
%%%%%%%%%%%%%%%%%%%%%

\appendix
\restartappendixnumbering
\section{UMS selection and field contamination}
    \label{app:UMS_selection}
    
    \begin{figure*}
        \centering
        \includegraphics[width = \linewidth]{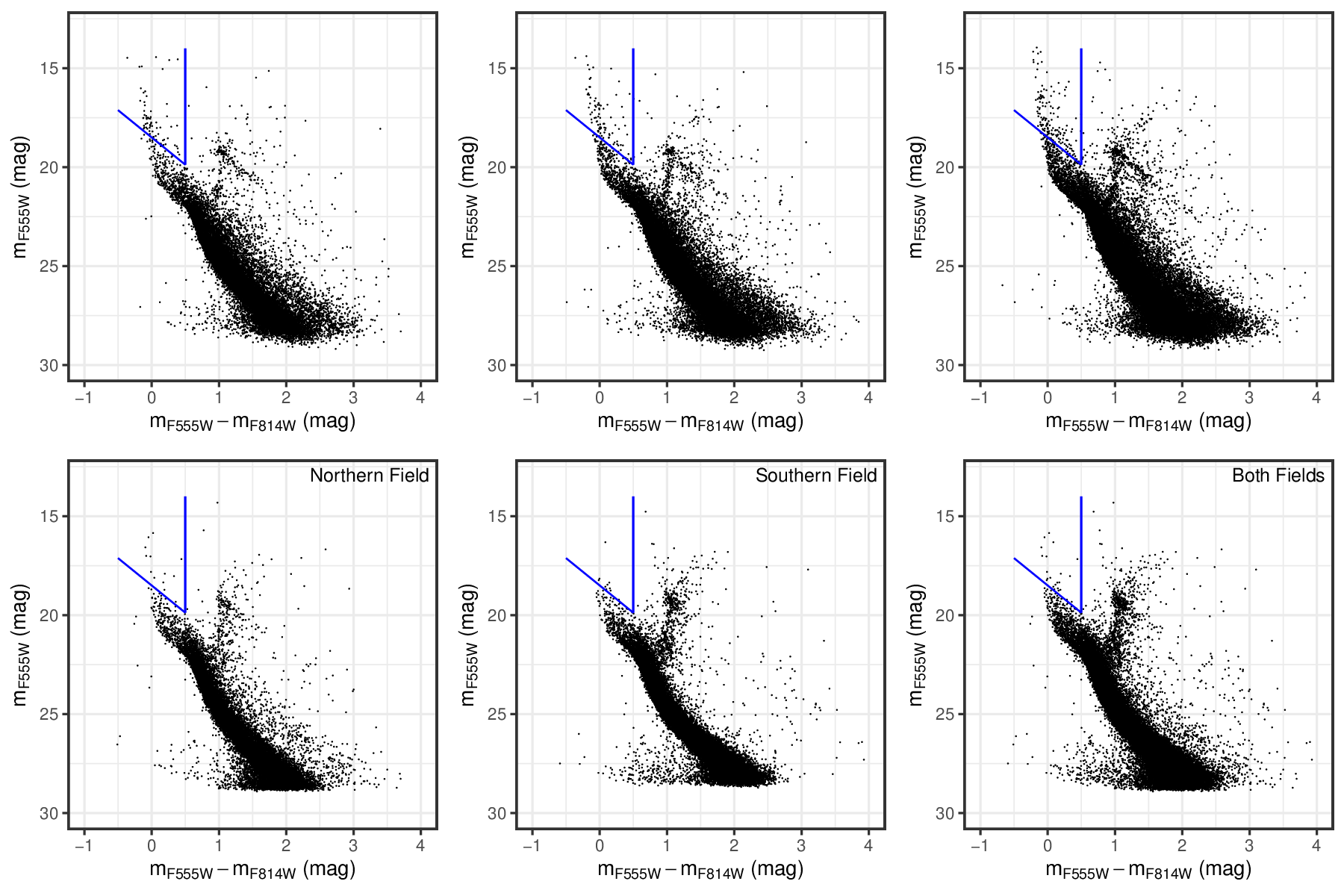}
        \caption{Optical CMDs of the main FoV of the MYSST survey subsampled to match the surface area of the northern (top left), southern (top center) and both reference fields (top right). For comparison the bottom row shows the actual CMDs of the northern (left), southern (center) and both reference fields (right). The blue lines indicate the limits of our UMS selection in all panels, demonstrating that we avoid most of the field contamination. }
        \label{fig:MYSST_main_subsampled}
    \end{figure*}
    
    In Section \ref{sec:UMS_ex} we select the brightest UMS sources in the MYSST FoV to estimate their individual stellar extinction and derive a reddening map for N44. While the massive stars in N44's star forming centers are young, the LMC field stars that contaminate the FoV of the survey are generally older evolved populations. Consequently, older stars of high enough mass in their post-main-sequence evolutionary phase can pollute the high brightness regime of the optical CMD. Therefore, we have to ensure that our UMS selection avoids as much field contamination as possible. 
    
    To do so we first limit the candidate UMS stars to objects bluer than 0.5 mag in $m_{F555W} - m_{F814W}$ to avoid the RGB and RC. As the lower brightness limit in $m_{F555W}$ we define a line parallel to the reddening vector. As mentioned in Section \ref{sec:Observations} the MYSST survey has also observed two LMC fields close to N44 for reference. By comparing the field CMDs to the main one we can determine the severity of the field contamination in relation to the chosen line and determine a suitable $m_{F555W}$-axis intercept. 
    
    To quantify the contamination of a given UMS selection we first sub-sample the main CMD as its FoV is much larger than the two fields. The latter have approximate surface areas of 1,650\,$\mathrm{pc^2}$ (northern) and 2,530\,$\mathrm{pc^2}$ (southern), while the main field covers about 33,440\,$\mathrm{pc^2}$. Using the area ratios we randomly sub-sample the main CMD once to match each field individually and once the combination of both fields. For a given UMS selection criterion we then count the selected stars in the sub-sampled main CMD and the respective field to determine the relative field contamination. The top row in Figure \ref{fig:MYSST_main_subsampled} shows examples for the sub-sampled main CMDs in comparison to the corresponding reference fields (bottom row). To account for randomness we repeat the sub-sampling procedure 5,000 times and average the results. 
    
    We determine a $m_{F555W}$-axis intercept of 18.5\,mag for our UMS selection criterion as the best compromise between selecting enough UMS sources to reasonably cover the main FoV and avoiding field contamination. Sub-sampling to the area and comparing to the CMD of the fields this selection criterion entails a $21.9 \pm 2.7\,\%$ contamination for the northern, $10.2 \pm 1.0\,\%$ for the southern and $14.8 \pm 1.1\,\%$ for both combined. Important to note here is that the 14 stars selected from the northern field could actually be UMS stars and not just old field contaminants. In the CMD of the northern field (bottom left panel, Figure \ref{fig:MYSST_main_subsampled}), we actually find a notable population of stars in the PMS region, contrary to the southern field CMD (bottom center panel), where this area is practically empty. Therefore, it is possible that the northern reference field might have captured a small star forming cluster and its UMS stars.
    
    Out of the 1,291 total sources selected by our criterion only 24 come from the reference fields. Given their CMD positions and the case we have made for the northern field we cannot easily dismiss these as non-UMS stars. Therefore, we decide to keep them as UMS candidates in our analysis and also derive (low resolution) extinction maps for the reference fields.   

\section{UMS extinction estimate error}
\restartappendixnumbering
\label{app:UMS_ex_error}

\begin{figure*}
    \centering
    \includegraphics[width = \linewidth]{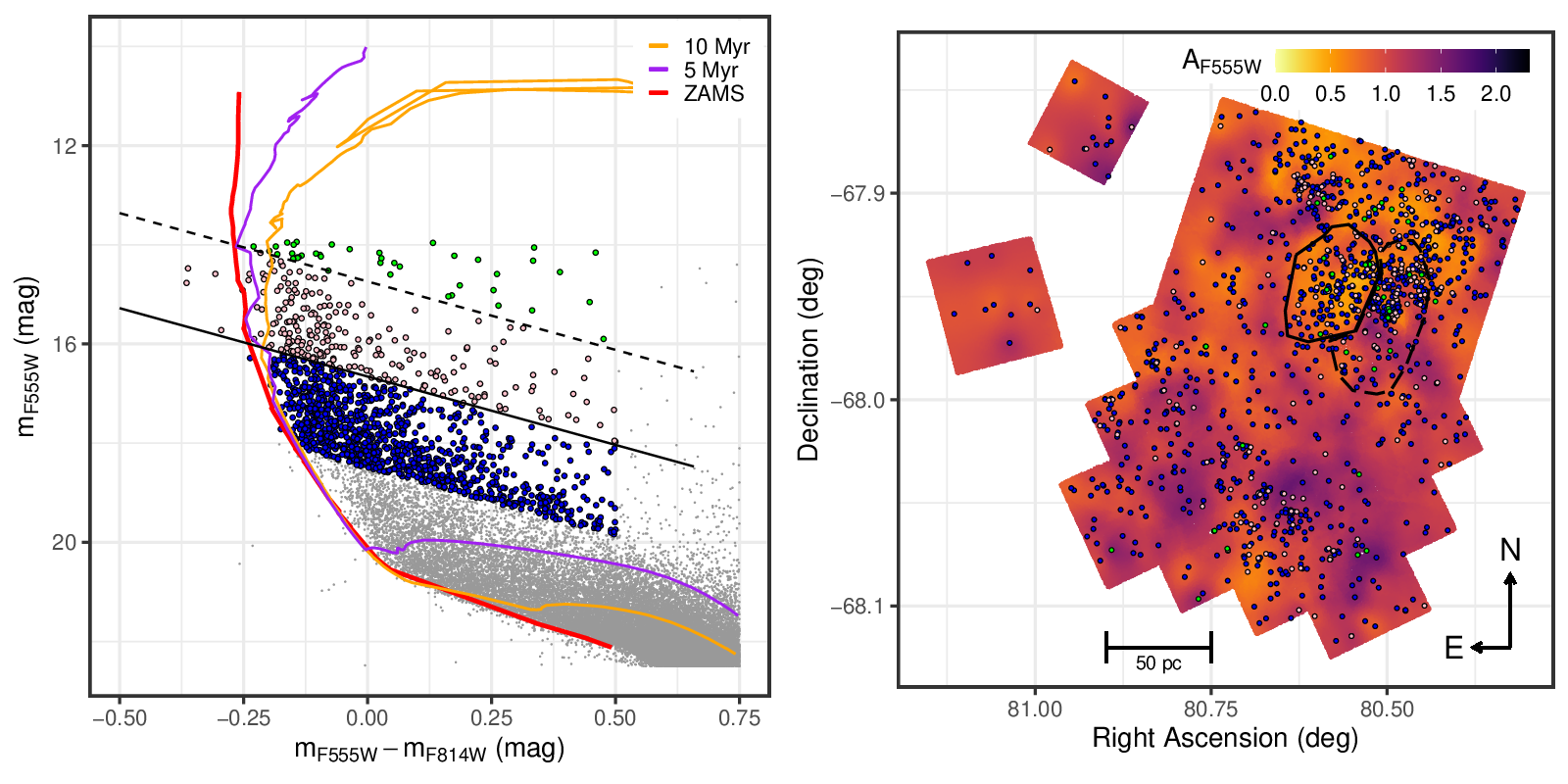}
    \caption{Zoom-in on the bright part of the MYSST CMD (left). The solid lines signify the ZAMS (red), a 5 Myr (purple) and a 10 Myr (orange) PARSEC isochrone, all corrected for the LMC distance modulus and MW foreground extinction. The large colored points mark our UMS selection as in the left panel of Figure \ref{fig:UMS_Selection}. The black solid and dashed lines show a projection of the points, where the 10 and 5 Myr isochrones start to move away from the ZAMS, along the reddening vector. The green points mark the UMS stars where the extinction measure would change if a 5 Myr isochrone is used instead of the ZAMS. If a 10 Myr isochrone is used instead of the ZAMS then the extinction measurement of the pink (+green) points is affected. The blue points are the UMS stars that are unaffected even in the 10 Myr case. Right: UMS extinction map of the MYSST survey (as in Figure \ref{fig:ExtinctionMap}). The large color points mark the position of the three groups of UMS stars identified in the left panel. Additionally, the solid and dashed black lines mark the position of the interior of N44's bubble and its western edge, respectively.}
    \label{fig:UMS_ex_err}
\end{figure*}

In our extinction estimation approach, presented in Section \ref{sec:UMS_ex}, we assume that our selected UMS sources should theoretically lie on the ZAMS. As already mentioned this ZAMS assumption does not necessarily hold for the rapidly evolving massive stars and their true position might actually differ from the ZAMS. Consequently, our approach may overestimate the extinction of some UMS stars, at worst providing only an upper extinction limit. We make the ZAMS assumption because we cannot easily date all UMS stars in our selection from the MYSST data alone and want our extinction measure to make the same systematic error everywhere. For some UMS stars in our sample, however, ages have been estimated in previous studies. E.g.~\cite{OeyMassey1995} find that the massive stars inside N44's bubble are about 10 Myr old, while the ones at the western bubble rim are younger at around 5 Myr. In this appendix we will briefly estimate the error in extinction that our approach entails for these two populations. 

To get a first idea of the systematic error of our extinction estimation approach, we estimate how many of our 1,291 UMS extinction probes would actually be affected if they were 5 or 10 Myr old instead of falling on the ZAMS. We do so by approximating the point on the ZAMS in the CMD where the 5 or 10 Myr PARSEC isochrone starts to significantly move away from the ZAMS track. We then project this point along the reddening vector to derive a threshold line in the CMD above which the UMS extinction measurement would change if the 5 or 10 Myr isochrone was used instead of the ZAMS track. For the 10 Myr isochrone we identify this point on the ZAMS at about 16 mag in F555W (including the LMC distance modulus and MW foreground extinction) and at 14 mag for the 5 Myr one. Note that the 5 Myr isochrone is fairly irregularly shaped, moving on and off the ZAMS, but appears to finally detach around 14 mag. The left panel in Figure \ref{fig:UMS_ex_err} shows these threshold lines in the MYSST CMD along with the corresponding isochrones and the UMS stars for which the extinction measurement would be affected by using the 5 or 10 Myr isochrone instead of the ZAMS. Here we find 41 UMS stars that would be affected if they were 5 Myrs old, and 276 in the 10 Myr case in total. Consequently, even if all UMS stars in our selection were 10 Myr old only 276 out of our 1,291 probes would even show a change in the measured extinction. As their spatial distribution in the right panel of Figure \ref{fig:UMS_ex_err} in comparison to the UMS extinction map shows, most of the affected UMS stars are located inside N44's bubble and its (western) rim. Particularly interesting in this diagram is that almost none of the affected UMS stars fall into the high-extinction "filament" extending from the southern bubble rim to the south-west corner of the FoV. As mentioned in the main text, this "filament" has no visible nebulous counterpart (i.e.~gas/dust) in long wavelength observations of N44 (e.g.~Spitzer), indicating that our approach is overestimating extinction in this region. If this is indeed the case, then Figure \ref{fig:UMS_ex_err} suggests that the UMS stars inside this "filament" must be even older than 10 Myrs. 

Lastly, to quantify the error of our approach, we measure extinction for the UMS stars in the bubble and rim with the correct 10 and 5 Myr isochrones instead of the ZAMS. In both cases we include all UMS stars that fall into the bubble (solid) and western rim (dashed) outlines in Figure \ref{fig:UMS_ex_err} and compute the median absolute error with respect to our ZAMS measurement. Here, we find $0.036_{-0.015}^{+0.026}\,\mathrm{mag}$ for the 241 UMS stars in the western rim and $0.045_{-0.007}^{+0.020}\,\mathrm{mag}$ for the 161 stars inside the bubble. Averaged across both populations the median absolute extinction error is $0.043_{-0.014}^{+0.019}\,\mathrm{mag}$ with a maximum of $0.27\,\mathrm{mag}$. The extinction error of our approach is, thus, overall fairly small inside and at the rim of the bubble. This does obviously not extend to UMS sources in our sample that are notably older than 10 Myr but at least confirms that our approach estimates the extinction around N44's super bubble fairly accurately and is at worst only an upper limit everywhere else. 

\section{UMS extinction and UMS broadening}
\restartappendixnumbering
\label{app:UMS_broadening}
    \begin{figure*}
        \centering
        \includegraphics[width = \linewidth]{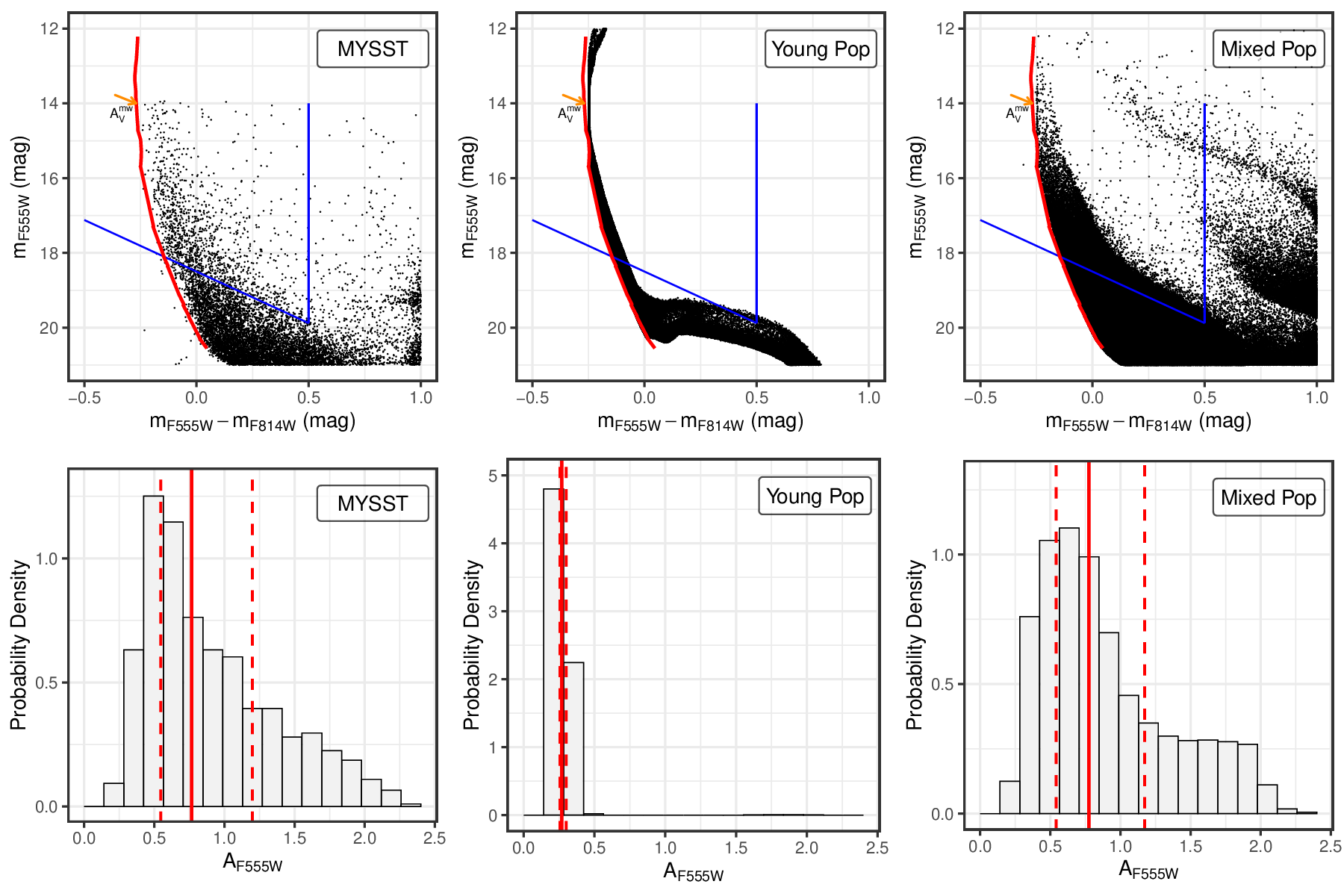}
        \caption{Zoom-in on the bright part of the CMD for the data from the MYSST survey (top left) and two synthetic data sets. One represents a single age population with a star formation episode between 5 and 5.6 Myr (top center), while the other emulates a mixed population with a constant star formation rate between 3.2 Myr and 12.6 Gyr (top right). Both synthetic populations are shifted according to the MW foreground reddening, but are intrinsically not affected by any extinction. In all three panels the blue lines indicate the limits of our UMS selection for the extinction estimation, while the red line marks the ZAMS, corrected for the LMC distance and MW foreground reddening. Lastly the orange arrow illustrates the shift of synthetic data and ZAMS due to the MW foreground reddening of $A_\mathrm{V}^\mathrm{mw} = 0.22\,$mag (or $A_\mathrm{F555W}^\mathrm{mw} = 0.223\,$mag). The bottom row shows the extinction distributions (including MW foreground) estimated by our method for the stars falling in the UMS selection of the corresponding data set in the top row. In all bottom panels the solid red line indicates the median estimated extinction, while the dashed red lines mark the 25\% and 75\% quantiles. }
        \label{fig:SynthPops_vs_MYSST}
    \end{figure*}
    
    In Appendix \ref{app:UMS_ex_error} we have discussed the error entailed in our extinction estimate due to assuming the ZAMS as the true position for the UMS stars. Beside differences in stellar age there are, however, other physical effects that result in a broadening of the UMS even in the absence of differential extinction. These include e.g.~unresolved binarity or metallicity gradients in the observed population. To ascertain the impact of these phenomena on our extinction estimation procedure we create two synthetic populations that are not affected by \textit{differential} reddening, such that all broadening of the UMS is caused by other effects. For both of the populations we assume an unresolved binarity fraction of 0.4 with a flat mass ratio distribution. We account for the LMC distance modulus ($(m-M)_0 = 18.55$) and include the constant shift due to MW foreground reddening ($A_\mathrm{V}^\mathrm{mw} = 0.22\,$mag). Lastly, we consider a metallicity spread of $[\mathrm{Fe}/\mathrm{H}] = -0.3$ to $-0.2$, corresponding to a range of $\mathrm{Z}=0.0076$ to $\mathrm{Z}=0.0096$ assuming the PARSEC solar metallicity of $\mathrm{Z_\odot}=0.01524$. 
    
    Our first synthetic data set represents an approximately single age population, formed with a constant star formation rate between 5 and 5.6\,Myr. The second one is a more extreme case, emulating a mixed age population resulting from a constant star formation rate between 3.2 Myr and 12.6 Gyr. The top panels in Figure \ref{fig:SynthPops_vs_MYSST} show these synthetic populations in the bright part of the CMD in comparison to the MYSST data. 
    
    We then repeat our UMS selection on these synthetic, intrinsically not reddened, populations and estimate "extinction" for the selected stars to quantify the impact of other broadening effects on our estimated extinction distribution. The results are displayed in the bottom panels of Figure \ref{fig:SynthPops_vs_MYSST} in comparison to the outcome on the MYSST data. 
    
    In the roughly single age synthetic population case we find a median "extinction" of $0.269_{-0.012}^{+0.032}\,$mag or $0.046_{-0.012}^{+0.032}\,$mag when subtracting the MW foreground. The error in the estimate introduced by non-extinction broadening effects is, therefore, fairly negligible when dealing with the ideal case of a single age population. In the case of the mixed age population, created by star formation with a constant rate over several tens of Gyrs, the outcome differs significantly. Here we find a median 'extinction' of $0.77_{-0.23}^{+0.40}\,$mag (or $0.56_{-0.23}^{+0.40}\,$mag subtracting $A_\mathrm{F555W}^\mathrm{mw}$) from the broadening of the UMS alone without any real differential extinction. This value is almost identical to our result on the real MYSST data and as we can see, comparing the bottom left and right panels of Figure \ref{fig:SynthPops_vs_MYSST}, the derived extinction distributions are similarly shaped too. While this result at first glance might call our MYSST extinction estimates into question, it is important to emphasize here that this synthetic mixed age population is not set up to match the LMC/N44 but as an extreme case to represent a worst case scenario. As previous studies \citep[e.g.][]{OeyMassey1995} indicate N44 is a region of multiple recent events of accelerated star formation rather than the outcome of a constant star formation process over several Gyrs. Additionally, our comparison with the observed LMC reference fields in Appendix \ref{app:UMS_selection} shows clearly that the background contamination in our UMS selection is fairly minimal. Consequently, the synthetic mixed age population is not likely to be a realistic representation of N44. Therefore, we conclude that the similarity of the MYSST extinction distribution and that of the mixed age synthetic population is a coincidence. Nevertheless, this experiment indicates that our extinction estimation procedure may be susceptible to larger systematic errors if applied to certain populations.

\section{}
\restartappendixnumbering

\begin{rotatetable*}
    \begin{deluxetable*}{lccccccccccccccccccccccc}
        \tabletypesize{\tiny}
        \centerwidetable
        \tablecaption{MYSST photometric catalog \label{tab:mysst_phot_cat}}
        \tablehead{
            \multicolumn{24}{c}{}\\
            & & & & & & \multicolumn{8}{c}{F555W} & \multicolumn{8}{c}{F814W} &  \\
            \colhead{ID} & \colhead{X} & \colhead{Y} & \colhead{R.A.} & \colhead{Decl.} & $\ldots$ & \colhead{type} & \colhead{$m$} & \colhead{$\sigma$} & \colhead{$\chi^2$} & \colhead{SNR} & \colhead{shrp} & \colhead{rnd} & \colhead{crwd} & \colhead{$f$} & \colhead{$m$} & \colhead{$\sigma$} & \colhead{$\chi^2$} & \colhead{SNR} & \colhead{shrp} & \colhead{rnd} & \colhead{crwd} & \colhead{$f$} & \colhead{$f_\mathrm{po}$} \\
            & (pixel) & (pixel) & (deg) & (deg) & & & (mag) & (mag) & & & & & (mag) & & (mag) & (mag) & & & & & (mag) & &
        }
        \startdata
        MYSST 052124.32-675620.82 & 27206.8 & 16934.1 & 80.35 & -67.94 & \multirow{20}{*}{$\ldots$} & 1 & 15.962 & 0.003 & 1.54 & 426 & -0.047 & 0.029 & 0.04 & 0 & 15.148 & 0.003 & 1.47 & 379 & -0.032 & 0.001 & 0.03 & 0 & 2\\
        MYSST 052140.51-675405.03 & 24943.0 & 20342.0 & 80.42 & -67.90 & & 1 & 16.804 & 0.005 & 1.38 & 220 & 0.009 & 0.032 & 0.10 & 0 & 14.762 & 0.002 & 0.97 & 461 & -0.007 & 0.008 & 0.07 & 0 & 2\\
        MYSST 052120.17-675548.08 & 27795.9 & 17749.0 & 80.33 & -67.93 & & 1 & 16.628 & 0.003 & 0.84 & 342 & -0.016 & 0.006 & 0.03 & 0 & 14.964 & 0.003 & 0.93 & 431 & -0.007 & -0.006 & 0.03 & 0 & 2\\
        MYSST 052207.85-675424.18 & 21085.5 & 19879.3 & 80.53 & -67.91 & & 1 & 16.889 & 0.004 & 1.02 & 291 & -0.017 & 0.014 & 0.06 & 0 & 14.749 & 0.003 & 1.09 & 430 & 0.010 & 0.012 & 0.04 & 0 & 2\\
        MYSST 052300.45-675319.41 & 13664.9 & 21508.7 & 80.75 & -67.89 & & 1 & 16.300 & 0.003 & 1.15 & 334 & 0.001 & 0.042 & 0.05 & 0 & 15.459 & 0.003 & 0.97 & 333 & -0.029 & 0.013 & 0.05 & 0 & 2\\
        MYSST 052152.65-675612.58 & 23217.2 & 17161.2 & 80.47 & -67.94 & & 1 & 15.735 & 0.002 & 0.91 & 534 & -0.019 & 0.001 & 0.02 & 0 & 15.852 & 0.004 & 1.06 & 269 & -0.033 & 0.012 & 0.02 & 0 & 2\\
        MYSST 052238.97-675509.11 & 16694.8 & 18764.8 & 80.66 & -67.92 & & 1 & 16.468 & 0.005 & 1.51 & 236 & -0.022 & 0.074 & 0.08 & 0 & 15.791 & 0.004 & 0.90 & 281 & -0.002 & 0.024 & 0.04 & 0 & 3\\
        MYSST 052157.45-675612.75 & 22541.9 & 17159.5 & 80.49 & -67.94 & & 1 & 15.927 & 0.002 & 1.03 & 468 & -0.030 & 0.004 & 0.04 & 0 & 15.872 & 0.005 & 1.37 & 206 & -0.030 & 0.021 & 0.05 & 0 & 2\\
        MYSST 052127.44-675437.94 & 26782.4 & 19508.9 & 80.36 & -67.91 & & 1 & 16.982 & 0.004 & 0.94 & 284 & 0.009 & 0.015 & 0.02 & 0 & 15.231 & 0.003 & 0.96 & 371 & 0.009 & 0.015 & 0.03 & 0 & 2\\
        MYSST 052151.66-675312.27 & 23376.1 & 21668.2 & 80.47 & -67.89 & & 1 & 16.090 & 0.007 & 2.43 & 167 & 0.012 & 0.034 & 0.22 & 0 & 15.834 & 0.007 & 1.78 & 149 & 0.030 & -0.026 & 0.18 & 0 & 3\\
        MYSST 052120.99-675526.03 & 27684.7 & 18301.3 & 80.34 & -67.92 & & 1 & 16.982 & 0.004 & 0.74 & 287 & -0.015 & 0.035 & 0.02 & 0 & 15.462 & 0.003 & 0.79 & 339 & -0.002 & -0.021 & 0.02 & 0 & 2\\
        MYSST 052151.10-675303.12 & 23456.9 & 21896.3 & 80.46 & -67.88 & & 1 & 15.832 & 0.003 & 1.22 & 398 & -0.003 & 0.042 & 0.07 & 0 & 15.903 & 0.004 & 1.10 & 242 & 0.010 & 0.017 & 0.07 & 0 & 2\\
        MYSST 052147.08-675608.33 & 24002.2 & 17263.8 & 80.45 & -67.94 & & 1 & 16.049 & 0.003 & 1.41 & 327 & -0.021 & 0.039 & 0.04 & 0 & 16.113 & 0.004 & 0.77 & 254 & -0.009 & 0.002 & 0.04 & 0 & 2\\
        MYSST 052214.31-675506.99 & 20170.8 & 18811.1 & 80.56 & -67.92 & & 1 & 17.208 & 0.003 & 1.25 & 406 & -0.021 & 0.020 & 0.04 & 2 & 15.067 & 0.002 & 1.16 & 661 & -0.008 & 0.014 & 0.02 & 2 & 2\\
        MYSST 052156.07-675530.81 & 22740.2 & 18207.3 & 80.48 & -67.93 & & 1 & 15.946 & 0.002 & 1.20 & 521 & 0.027 & -0.010 & 0.05 & 2 & 16.098 & 0.004 & 1.33 & 246 & 0.013 & 0.010 & 0.05 & 2 & 2\\
        MYSST 052136.23-675334.68 & 25551.0 & 21097.1 & 80.40 & -67.89 & & 1 & 16.224 & 0.005 & 1.72 & 221 & -0.034 & 0.021 & 0.05 & 0 & 16.112 & 0.005 & 1.00 & 226 & -0.021 & 0.007 & 0.04 & 0 & 2\\
        MYSST 052230.62-675327.57 & 17875.2 & 21302.2 & 80.63 & -67.89 & & 1 & 15.987 & 0.003 & 1.17 & 409 & -0.023 & 0.003 & 0.01 & 0 & 16.144 & 0.004 & 0.91 & 249 & -0.018 & 0.001 & 0.01 & 0 & 3\\
        MYSST 052218.45-675327.24 & 19593.2 & 21306.9 & 80.58 & -67.89 & & 1 & 16.065 & 0.003 & 1.13 & 403 & -0.008 & 0.012 & 0.03 & 0 & 16.119 & 0.004 & 0.91 & 251 & 0.009 & 0.015 & 0.05 & 0 & 2\\
        MYSST 052147.82-675521.85 & 23903.3 & 18425.9 & 80.45 & -67.92 & & 1 & 17.120 & 0.003 & 1.03 & 360 & -0.028 & 0.023 & 0.02 & 0 & 15.727 & 0.003 & 1.17 & 357 & -0.013 & 0.013 & 0.02 & 0 & 2\\
        MYSST 052229.28-675354.70 & 18063.5 & 20622.9 & 80.62 & -67.90 & & 1 & 16.152 & 0.004 & 1.37 & 309 & -0.020 & 0.027 & 0.02 & 0 & 16.233 & 0.005 & 0.90 & 234 & -0.006 & -0.014 & 0.01 & 0 & 2\\
        \enddata
        \tablecomments{The omitted ($\ldots$) columns list combined values (over the two filters) for the PSF fit quality $\chi^2$, signal to noise ratio \textit{SNR}, sharpness \textit{shrp}, roundness \textit{rnd}, crowding \textit{crwd} and direction of major axis (if not round) \textit{mjaxdir}.}
        \end{deluxetable*}    
    \end{rotatetable*} 

\end{document}